\documentclass[aps,prd,twocolumn,superscriptaddress,floatfix,amsmath,amssymb,amsfonts,nofootinbib,10pt]{revtex4-2}

\usepackage[utf8]{inputenc}
\usepackage[T1]{fontenc}
\usepackage{lmodern}
\usepackage{bm}
\usepackage{hyperref}
\usepackage{graphicx}
\usepackage[breakwords]{truncate}
\usepackage{amsmath}
\usepackage{mathrsfs}
\usepackage{subcaption}
\usepackage{tikz}
\usepackage{amssymb}
\usepackage{ragged2e}
\usepackage[font={small}]{caption}

\usepackage{dsfont}
\usepackage{amssymb,amsthm}  
\usepackage{mathtools}
\newcommand{\ii}{\mathrm{i}}
\usepackage{physics}
\usepackage{graphicx,color}

\usepackage[shortlabels]{enumitem}
\usepackage[normalem]{ulem}

\begin{document}

\title{Phase space analysis of Bell inequalities for mixed Gaussian states}

\author{Gurpahul Singh}
\email{gsingh@perimeterinstitute.ca}
\affiliation{Perimeter Institute for Theoretical Physics, Waterloo, Ontario, N2L 2Y5, Canada}
\affiliation{Department of Physics and Astronomy, University of Waterloo, Waterloo, Ontario N2L 3G1, Canada}
\author{Kelly Wurtz}
\email{kwurtz@uwaterloo.ca}
\affiliation{Perimeter Institute for Theoretical Physics, Waterloo, Ontario, N2L 2Y5, Canada}
\affiliation{Institute for Quantum Computing, University of Waterloo, Waterloo, Ontario, N2L 3G1, Canada}
\affiliation{Department of Applied Mathematics, University of Waterloo, Waterloo, Ontario, N2L 3G1, Canada}
\author{Eduardo Mart\'in-Mart\'inez}
\email{emartinmartinez@uwaterloo.ca}
\affiliation{Perimeter Institute for Theoretical Physics, Waterloo, Ontario, N2L 2Y5, Canada}
\affiliation{Institute for Quantum Computing, University of Waterloo, Waterloo, Ontario, N2L 3G1, Canada}
\affiliation{Department of Applied Mathematics, University of Waterloo, Waterloo, Ontario, N2L 3G1, Canada}
\date{\today}

\begin{abstract}
We present a phase space formalism to evaluate Bell inequality violations in continuous variable systems. By doing so we can generalize previous analyses (which have dealt only with pure states) to arbitrary mixed states. We leverage these results to analyze the effect of temperature on violations of Bell inequalities in a two-mode squeezed thermal state, which can become useful in tests of local realism in the presence of thermal noise. We also explore the non-monotonic relationship between the violations of Bell inequalities and the amount of entanglement present in this family of mixed states. Additionally, we discuss the optimal choices of pseudospin operators for states beyond the two-mode squeezed vacuum. 

\end{abstract}

\maketitle

\section{Introduction}
\label{sec:intro}
The violation of Bell inequalities demonstrates that entangled systems can defy local realism and exhibit non-classical behavior. The original Bell inequality \cite{Bell_paper} was formulated for a pair of entangled spin-1/2 particles, with operators acting on a finite-dimensional Hilbert space. However, the EPR paradox \cite{EPR_paper} was formulated with a wavefunction expressed in an infinite-dimensional Hilbert space. Because the Wigner function \cite{Wigner_first_paper} of this state is positive everywhere in phase space, reminiscent of a classical probability distribution, Bell erroneously believed that this state could not violate local realism \cite{Bell_positive_Wigner}. 

It was later established that the EPR state is a special case of a two-mode squeezed (TMS) vacuum state \cite{SMS1,SMS2} --- a pure entangled state of two harmonic oscillators, which reduces to the EPR state in the limit of infinite squeezing \cite{EPR_continuous_experiment2}. Subsequently, the TMS vacuum state was shown to exhibit Bell inequalities violation (BIV) both theoretically \cite{Nonlocality_Wigner1, Nonlocality_Wigner2, Nonlocality_Wigner3, Chen_pseudospin, Gour_position_rep} and experimentally \cite{Violation_Bell_Gaussian1, Violation_Bell_Gaussian2, Experiment_violation2, Experiment_violation_space}. One way of observing BIV in continuous variable systems such as a TMS state is to define \emph{pseudospin} operators on the infinite-dimensional Hilbert space, in analogy to the Pauli operators in spin-1/2 systems. The Bell inequality, constructed from these pseudospin operators, then serves as a way to test non-classicality in continuous variable systems.

Understanding what influences BIV in continuous variable systems is crucial not only for foundational physics but also for practical applications in quantum information science. A number of protocols in quantum key distribution \cite{Usecase1} and quantum random number generation \cite{usecase2} rely on BIV to ensure that the relevant systems exhibit non-classical correlations. The past two decades have witnessed significant efforts to generalize the analysis of BIV in the TMS vacuum state. These efforts have identified dependence on the squeezing angle \cite{Main_paper}, measurement angles, and choice of pseudospin operators \cite{Gour_position_rep, Larsson_fock_basis_grouping, Main_paper2}. Despite this progress, little is known about states outside the TMS vacuum state. One very useful departure from previous literature is to consider the effect of thermality, as temperature is one of the most significant sources of decoherence in technological applications of entanglement. Indeed, it has been previously shown in qubit systems \cite{Bell_violation_thermal1, Bell_violation_thermal2, Bell_violation_thermal3} that thermality degrades entanglement and reduces BIV. For continuous variable systems, however, the extent to which temperature affects BIV has yet to be studied. We begin this analysis by examining BIV in TMS thermal states. 

A TMS thermal state is a mixed state, which 
complicates the analysis in the usual Hilbert space picture. However, we may exploit the fact that the TMS thermal state, like the TMS vacuum, is a Gaussian state, and thus can be fully represented as a Gaussian distribution in a four-dimensional phase space. If we could also represent the pseudospin operators in phase space, we would immensely simplify the BIV to any Gaussian state of a continuous variable system.

In this paper, we find the Wigner-Weyl (phase space) representation of a particular class of pseudospin operators, introduced originally in \cite{Main_paper2}. Remarkably, while it was proven that other choices of pseudospin operators yielded stronger BIV for the TMS vacuum, using the phase space formalism we show that the operators in \cite{Main_paper2} actually outperform those choices when thermality is involved. We also discuss the question of optimality of BIV for the choice of pseudospin operators when working with their phase space representation. Finally, we show, perhaps unintuitively, that the strength of the BIV in the TMS thermal state display a non-monotonic behaviour with the total amount of entanglement in the state.


The organization of the paper is as follows. Section \ref{section:operator_presentation} introduces the CHSH inequality for continuous variable systems and presents the class of pseudospin operators proposed in~\cite{Main_paper2}. In Sec.~\ref{sec:Bell_op_TMST} we outline the derivation of the Wigner representation of the pseudospin operators, then use it to calculate their correlation functions for a TMS thermal state, and analyze how the CHSH violation varies with temperature, squeezing, and choice of operators. In Sec.~\ref{sec:entanglement}, we look at the relationship between entanglement and strength of CHSH violation for the TMS thermal states, focusing on the non-monotonic relation between them. Finally, in Sec.~\ref{sec:operator_choice}, we discuss why optimal position-space operators are currently absent from the literature, and how this research opens a new avenue to discuss optimization of Bell operators in continuous variable systems.

\section{Bell inequalities in continuous variable systems}\label{section:operator_presentation}
In this section, we will provide an overview of the pseudospin operator approach to evaluating Bell inequalities. 

\subsection{The CHSH inequality}

The CHSH inequality \cite{CHSH_Inequality} simplifies Bell's theorem for the case that each observer can perform two measurements, each with two possible outcomes. The inequality is a statement bounding the expectation value of the Bell operator:
\begin{equation}
\label{eq:Bell_operator}
\hat{B} \coloneqq \hat{\sigma}_{\bm{n}_1}^{(1)} \hat{\sigma}_{\bm{n}_2}^{(2)}  + \hat{\sigma}_{\bm{n}_1}^{(1)}  \hat{\sigma}_{\bm{n}_2'}^{(2)} +  \hat{\sigma}_{\bm{n}_1'}^{(1)} \hat{\sigma}_{\bm{n}_2}^{(2)} -\hat{\sigma}_{\bm{n}_1'}^{(1)} \hat{\sigma}_{\bm{n}_2'}^{(2)},
\end{equation}
where
\begin{equation}
    \hat{\sigma}_{\bm{n}_1}^{(1)} \hat{\sigma}_{\bm{n}_2}^{(2)} \equiv ( \bm{n}_1 \cdot \hat{\bm{\sigma}}_1 ) \otimes ( \bm{n}_2 \cdot \hat{\bm{\sigma}}_2 ),
\end{equation}
and where 1 and 2 label the subsystems of a bipartite system. Traditionally, $\hat{\bm{\sigma}}_i = ( \hat{\sigma}_x^{(i)}, \hat{\sigma}_y^{(i)}, \hat{\sigma}_z^{(i)} )$ is taken to be the vector of Pauli spin operators acting on subsystem $i$, however $\hat{\bm{\sigma}}_i$ can be a vector of any set of dichotomous observables (those which define a measurement with a binary outcome). The unit vectors $\bm{n}_i$ are defined as
\begin{equation}
\bm{n}_i = \left( \sin{\theta_i}\cos{\phi_i}, \sin{\theta_i}\sin{\phi_i}, \cos{\theta_i} \right),
\label{eq:n_vector}
\end{equation}
such that $\hat{\sigma}_{\bm{n}_i} = \bm n_i \cdot \hat{\bm \sigma}_i$ is an operator in the direction ${\bm{n}_i}$ in the space of operators spanned by $\hat{\bm{\sigma}}_i$.

The Bell operator $\hat{B}$ can be used to classify states that do not admit a local realistic description. Defining the following correlation function for a state $\hat{\rho}$ as
\begin{equation} 
\label{eq:correlation_fn} 
P( \bm{n}_1, \bm{n}_2 ) \coloneqq \langle\hat{\sigma}^{(1)}_{\bm{n}_1}\hat{\sigma}^{(2)}_{\bm{n}_2}\rangle_{\hat{\rho}},
\end{equation} 
the CHSH inequality asserts that any local realistic theory must satisfy
\begin{align}
\label{eq:Bellinequality}
| \langle \hat{B} \rangle_{\hat{\rho}} | &= | P( \hat{n}_1, \hat{n}_2 ) + P( \hat{n}_1, \hat{n}_2' ) + P( \hat{n}_1', \hat{n}_2 ) - P( \hat{n}_1', \hat{n}_2' ) | \nonumber \\
&\leq 2.
\end{align}
In quantum mechanical systems, the expectation value of this operator is at most $2\sqrt{2}$. This bound is known as Tsirelson's bound \cite{Cirelsonbound}. Thus, in quantum mechanics, violations of local realism happen for
\begin{equation}
    2 < |\langle \hat{B} \rangle_{\hat{\rho}} | \leq 2\sqrt{2}.
\end{equation}
With a certain choice of $\hat\sigma_{\bm n_i}$, Tsirelson's bound can be saturated for any maximally entangled state.

\subsection{Pseudospin operators for the two-mode squeezed state}\label{subsec:binned_pos_op}
Evaluating $\langle \hat{B}\rangle_{\hat{\rho}}$ for continuous variable systems requires a set of three (infinite-dimensional) dichotomous observables. The typical approach is to build three ``pseudospin'' operators. To construct the first, we partition the eigenvectors of an observable of the system into two distinct subsets (which need not be connected) and build a dichotomous operator that has a degenerate spectrum with an eigenvalue of $+1$ for all the eigenstates in one subset and $-1$ to the other. The other two operators are chosen such that they, together with the first operator, form an $\mathfrak{su}(2)$ algebra. In a qubit system, choosing one spin operator uniquely defines the other two (modulo Bloch sphere rotations). However, in the case of an infinite dimensional system, given a pseudospin operator, there are infinitely many choices for the other two that form an $\mathfrak{su}(2)$ algebra which are not equivalent under Bloch sphere rotations. This freedom makes finding optimal pseudospin operators for a given entangled state a challenging computational problem. 

In particular, we would like to determine a set of operators that maximizes violations of the CHSH inequality for a physically realizable entangled state in a continuous variable system. For example, a ubiquitous bipartite continuous variable entangled state is the TMS vacuum state: 
\begin{align}\label{eq:TMSV_state}
    |\text{TMSV}\rangle = \frac{1}{\cosh{r}}\sum_{n=0}^{\infty} e^{-2\ii n \phi} \tanh^n{r} \ket{n}\ket{n}\,,
\end{align}
where $r$ is the squeezing parameter and $\phi$ represents the squeezing angle. This is a particularly simple pure entangled state whose simplicity was exploited to find a particular set of pseudospin operators \cite{Chen_pseudospin} that optimize BIV for the TMS vacuum state at any level of squeezing \cite{Gour_position_rep}. 
The expressions for these pseudospin operators have been included in Appx.~\ref{appx:pseudospin_ops}.

Our objective is to build an efficient framework to analyze CHSH violations beyond the TMS vacuum for any Gaussian state, mixed or pure. To do this, we will formulate the problem in phase space and use Gaussian quantum mechanics. With this in place, we will then analyze TMS thermal state in detail. 

For a phase space analysis of the problem, we need a set of pseudospin operators in a continuous basis representation, such as the position representation. While the set of operators in \cite{Chen_pseudospin} is optimal for the TMS vacuum state, it was presented exclusively in the Fock basis, which does not directly lend itself to a phase-space formulation of the problem. A possible approach would be to convert these optimal operators for the TMS vacuum from the Fock basis to the position basis. However, the conversion is highly nontrivial, and, more importantly, it is unclear whether a set of operators that are optimal for the vacuum would also be optimal operators for any other state. Discussions about this transformation and finding optimal position-space operators will be the subject of Sec.~\ref{sec:operator_choice}.

Another construction of pseudospin operators built from a bipartition of the eigenvectors of the position operator was given in~\cite{Gour_position_rep}. There, the position space is split into two bins of $q\ge0$ and $q<0$. In this construction, the $\hat \sigma_z$ pseudospin operator is chosen to be the sign function of the position operator, such that the eigenvalue is 1 for eigenstates with positive position and -1 for eigenstates with negative position. The operators are given  explicitly as:
\begin{subequations}
\label{eq:pseudo_pos}
    \begin{align}
    \hat{\Pi}_z &= -\int_{-\infty}^{\infty}\dd{q}\ket{q}\!\bra{-q}\\ 
    \hat{\Pi}_x &=\int_0^{\infty} \dd{q}\left\{\ket{q}\!\bra{q}-\ket{-q}\!\bra{-q}\right\} \\
    \hat{\Pi}_y &=-\ii\int_0^{\infty} \dd{q}\left\{\ket{q}\!\bra{-q}-\ket{-q}\!\bra{q}\right\}\,.
\end{align}
\end{subequations}

While natural from a physical point of view, this construction does not yield optimal violations of the CHSH inequality for the TMS vacuum state \cite{Gour_position_rep}. 

Ref.~\cite{Main_paper2} presents a larger class of similar pseudospin operators
in which the position space is partitioned into
an infinite number of intervals $[nl, (n+1)l]$ of length $l$. The pseudospin operator $\hat{S}_z$ is defined such that it assigns an eigenvalue $+1$ to position states in intervals for which $n$ is even, and $-1$ where $n$ is odd, effectively binning the values of the position into a binary class\footnote{A similar binning strategy, grouping different eigenstates of the number operator into two bins, has also been used to build pseudospin operators~\cite{Larsson_fock_basis_grouping}. For more details see Appx.~\ref{appx:pseudospin_ops}.}. The parameter $l$ introduces additional freedom that can be exploited for optimization of the CHSH violation. We will be primarily working with these operators henceforth and will discuss later in more detail the advantages and shortcomings of this choice.

Concretely, the binned operator $\hat{S}_z(l)$ is defined as
\begin{equation}
    \label{eq:Sz}
    \hat{S}_z(l) \coloneqq \sum_{-\infty}^{\infty}(-1)^n\!\!\!\! \int_{nl}\limits^{(n+1)l} \!\!\!\dd{q} \ket{q}\!\bra{q}.
\end{equation}
We will require that the operators satisfy the $\mathfrak{su}(2)$ algebra. Once $\hat{S}_z$ is defined, $\hat{S}_x$ and $\hat{S}_y$ are then determined up to a rotation around the $z$ axis. The optimal choice (see Sec.~\ref{sec:operator_choice}) is 
\begin{align}
    \label{eq:Sx}
    \hat{S}_x(l) &= \hat{S}_{+}(l) + \hat{S}_{-}(l) \\ \label{eq:Sy}
    \hat{S}_y(l) &= -\ii\left(\hat{S}_{+}(l) - \hat{S}_{-}(l)\right)\,, 
\end{align}
where 
\begin{align}\label{eq:S+}
    \hat{S}_+(l) &= \sum_{n=-\infty}^{\infty} \int_{2nl}\limits^{(2n+1)l}\!\!\!\dd{q}\ket{q}\!\bra{q+l}\\ \label{eq:S-}
    \hat{S}_-(l) &= \sum_{n=-\infty}^{\infty} \int_{(2n+1)l}\limits^{(2n+2)l}\!\!\!\dd{q}\ket{q}\!\bra{q-l}\,,
\end{align}
noting that $\hat{S}_+^{\dagger}(l) = \hat{S}_-(l)$.  Like the Pauli operators, $\hat{S}_x$, $\hat{S}_y$, and $\hat{S}_z$ have eigenvalues $\pm 1$ and square to the identity.

We will now express the Bell operator \eqref{eq:Bell_operator} in terms of the binned pseudospin operators. Without loss of generality, we can set all azimuthal angles $\phi_i$ in both $\bm{n}_1$ and $\bm{n}_2$ to 0, since the Bell operator is only sensitive to the relative angles between the measurement settings. The correlation function \eqref{eq:correlation_fn} for the TMS state  then takes the form 
\begin{align}
    P(l,\theta_1,\theta_2) &= \sin{\theta_1}\sin{\theta_2}\langle\hat{S}^{(1)}_x(l)\hat{S}^{(2)}_x(l)\rangle_{\hat{\rho}} \nonumber \\
    &\qquad \quad +\cos{\theta_1}\cos{\theta_2}\langle\hat{S}^{(1)}_z(l)\hat{S}^{(2)}_z(l)\rangle_{\hat{\rho}}.
\end{align}
Since we have set the azimuthal angles to zero, $\langle\hat{S}^{(1)}_y(l)\hat{S}^{(2)}_y(l)\rangle_{\hat{\rho}}$~does not contribute to the correlation function. In principle, cross-terms such as $\langle\hat{S}^{(1)}_z(l)\hat{S}^{(2)}_x(l)\rangle_{\hat{\rho}}$ could appear, but these vanish for the TMS vacuum \cite{Main_paper} and thermal state (see Appx.~\ref{Appendix:Correlation_funcs_TMST}).

Optimizing $\langle\hat{B}\rangle_{\hat{\rho}}$ with respect to the measurement angles \cite{Main_paper} shows that the optimal choice is $\theta_1 = 0$, $\theta_1' = \pi/2$, and $\theta_2 = -\theta_2'$, with 
\begin{equation}\label{eq:or_par}
    \theta_2 = \arctan\left[\frac{\langle\hat{S}^{(1)}_x(l)\hat{S}^{(2)}_x(l)\rangle_{\hat{\rho}}}{\langle\hat{S}^{(1)}_z(l)\hat{S}^{(2)}_z(l)\rangle_{\hat{\rho}}}\right]\,.
\end{equation}
With these angles, the expectation value of the Bell operator for the TMS thermal state $\hat{\rho}_{\text{ST}}$ becomes
\begin{equation}\label{eq:Bell_op}
    \langle\hat{B}\rangle_{\hat{\rho}_{\text{ST}}} = 2\sqrt{\langle\hat{S}^{(1)}_z(l)\hat{S}^{(2)}_z(l)\rangle_{\hat{\rho}_{\text{ST}}}^2+\langle\hat{S}^{(1)}_x(l)\hat{S}^{(2)}_x(l)\rangle_{\hat{\rho}_{\text{ST}}}^2}\,.
\end{equation}
Note that the above equation is valid for the TMS vacuum state $\hat{\rho}_{\text{SV}}$ as well. In Sec.~\ref{sec:Bell_op_TMST}, we will evaluate the Bell operator expectation value via the above correlation functions for the TMS thermal state.

\section{Bell's inequality for a TMS Thermal State} \label{sec:Bell_op_TMST}

Since the TMS thermal state is a Gaussian state, it is most convenient to bypass any Hilbert space calculations and instead express the pseudospin operators in the phase-space (Wigner-Weyl) representation.


\subsection{Pseudospin correlation functions from phase space operators}
\label{subsec:correlation_funcs_Wigner}
We first obtain the Wigner-Weyl representation of the pseudospin operators given in Eqs.~\eqref{eq:Sz}--\eqref{eq:Sy}. The Wigner-Weyl representation $O(q,p)$ of an operator $\hat{O}$ is (see, for instance, \cite{Hall2013, GROENEWOLD1946405, Eduardo_lecture_notes})
\begin{equation}\label{eq:Wig_trans}
    O(q,p) = \frac{1}{2\pi}\int_{-\infty}^{\infty}\!\!\!\dd{x}e^{-\ii px}\mel{q+\tfrac{1}{2}x}{\hat{O}}{q-\tfrac{1}{2}x}.
\end{equation} 
We will be using natural units ($
=1, k_B=1$) throughout the paper. Applying Eq.~\eqref{eq:Wig_trans} to $\hat{S}^{(i)}_z(l)$, where $i \in~\{ 1, 2\}$ denotes the subsystem, we find that the Wigner-Weyl representation is given (see Appx.~\ref{appx:correlation_funcs}) by
\begin{equation}\label{eq:Sz_Wig_trans}
    \!\!\!S_z^{(i)}(l, q_i,p_i) = \!\!\! \sum_{n=-\infty}^{\infty}\! (-1)^n \!\!\!\!\! \int_{nl}\limits^{(n+1)l} \!\!\!\! \dd{q} e^{-2\ii p_i(q-q_i)} \delta(q-q_i).
\end{equation}
The expectation value of any operator $\hat{O}$ for a state $\hat{\rho}$ described by the Wigner function $W_{\hat{\rho}}(q,p)$ can be simply expressed as 
\begin{equation}
\label{eq:expectation_value_Wigner}
    \langle\hat{O}\rangle_{\hat{\rho}} = \iint\dd{q}\dd{p} W_{\hat{\rho}}(q,p) O(q,p). 
\end{equation}
Using \eqref{eq:expectation_value_Wigner}, we find the correlation function $\langle\hat{S}^{(1)}_z(l)\hat{S}^{(2)}_z(l)\rangle_{\hat{\rho}}$ to be
\begin{multline}
\label{eq:correlationSzSz}
\langle\hat{S}^{(1)}_z(l)\hat{S}^{(2)}_z(l)\rangle_{\hat{\rho}} = 
\sum_{m,n=-\infty}^{\infty}(-1)^{m+n} \int_{-\infty}^{\infty}\!\!\!\dd{p_1}\dd{p_2} \\  \times \int_{nl}\limits^{(n+1)l}\!\!\!\!\!\! \dd{q_1} \int_{ml}\limits^{(m+1)l}\!\!\!\!\!\! \dd{q_2} W_{\hat{\rho}}(q_1,q_2,p_1,p_2).
\end{multline}
Similarly, for $\langle\hat{S}^{(1)}_x(l)\hat{S}^{(2)}_x(l)\rangle_{\hat{\rho}}$, the correlation function becomes 
\begin{align}
\label{eq:correlationSxSx}
\langle\hat{S}^{(1)}_x(l)\hat{S}^{(2)}_x(l)\rangle_{\hat{\rho}} &= 2\Re\{\langle\hat{S}^{(1)}_+(l)\hat{S}^{(2)}_+(l)\rangle_{\hat{\rho}}\nonumber \\ &\phantom{={}}+ \langle\hat{S}^{(1)}_+(l)\hat{S}^{(2)}_-(l)\rangle_{\hat{\rho}}\},
\end{align}
where (see Appx.~\ref{appx:correlation_funcs})
\begin{align}\label{eq:correlationS+S+}
    \langle\hat{S}^{(1)}_+(l)\hat{S}^{(2)}_+(l)\rangle_{\hat{\rho}} = 
    \sum_{m,n=-\infty}^{\infty} \int_{-\infty}^{\infty}\!\!\!\dd{p_1} \dd{p_2}\!\!\! \int_{2nl}\limits^{(2n+1)l}\!\!\!\! \!\!\dd{q_1}\nonumber\\ \times \int_{2ml}\limits^{(2m+1)l} \!\!\!\!\!\!\!\dd{q_2} \,
     e^{\ii(p_1+p_2)l}\, W
_{\hat{\rho}}(q_1+\tfrac{l}{2},q_2+\tfrac{l}{2},p_1,p_2),\\
\label{eq:correlationS+S-}
    \langle\hat{S}^{(1)}_+(l)\hat{S}^{(2)}_-(l)\rangle_{\hat{\rho}} = 
    \sum_{m,n=-\infty}^{\infty} \int_{-\infty}^{\infty}\!\!\!\dd{p_1}\dd{p_2}\!\!\!\! \int_{2nl}\limits^{(2n+1)l}\!\!\!\!\!\! \dd{q_1}\nonumber\\ \times \int_{(2m+1)l}\limits^{(2m+2)l}\!\!\!\!\!\! \dd{q_2}
    e^{\ii(p_1-p_2)l}\, W_{\hat{\rho}}(q_1+\tfrac{l}{2},q_2-\tfrac{l}{2},p_1,p_2)\,.
\end{align}
In the next section, we will find these expectation values specifically for the TMS thermal state.

\subsection{Pseudospin correlation functions for a TMS thermal state}\label{subsec:correlation_funcs_TMST}

The Wigner function for the TMS thermal state $\hat{\rho}_{\text{ST}}$ can be obtained directly from its covariance matrix, as detailed in Appx.~\ref{appx:cov_and_Wig}. We find
%
%
\begin{multline}
    \label{eq:Wigner_TMS_thermal}
    W_{\hat{\rho}_{\text{ST}}}(q_1,q_2,p_1,p_2) = \\
    \frac{1}{(\nu(T)\pi)^2}e^{\frac{1}{\nu(T)}[2\sinh(2r)(q_1q_2-p_1p_2) - \cosh{(2r)(q_1^2+q_2^2+p_1^2+p_2^2)} ]}, \raisetag{14mm}
\end{multline}
where $\nu(T) = \coth{(\omega/2T)}$. As $T \rightarrow 0$ and $\nu(T) \rightarrow 1$, we recover the Wigner function of the TMS vacuum state~\cite{Wigner_function_TMSV}.

Using this Wigner function, we can evaluate the pseudospin correlation functions \eqref{eq:correlationSzSz} and \eqref{eq:correlationSxSx}. After integrating over $p_1$ and $p_2$ we obtain
\begin{equation}\label{eq:correlationSzSz_TMST} \langle\hat{S}^{(1)}_z(l)\hat{S}^{(2)}_z(l)\rangle_{\hat{\rho}_{\text{ST}}} = \frac{1}{\pi}\sum_{m,n=-\infty}^{\infty}(-1)^{m+n} Z_{n,m}\,,
\end{equation}
where 
\begin{multline}\label{eq:Znm}
        Z_{n,m} = \frac{l}{(\nu(T))^{3/2}}\sqrt{\frac{\pi}{\gamma_2}} \int_0^1 \dd{z} e^{-\gamma_1(z+n+m)^2\tfrac{l^2}{4}} \\
        \times \left\{\erf[\tfrac{l}{2}\sqrt{\gamma_2}(z+n-m)]+ \erf[\tfrac{l}{2}\sqrt{\gamma_2}(z-n+m)]\right\},  \raisetag{14mm}
\end{multline}
with
\begin{equation}
\gamma_1 = \gamma_1(r,T) = \tfrac{2}{\nu(T)}e^{-2r},
\gamma_2 = \gamma_2(r,T)= \tfrac{2}{\nu(T)}e^{2r}.
\end{equation}
We will evaluate Eq.~\eqref{eq:Znm} numerically. 

Performing an analogous computation to find $\langle\hat{S}^{(1)}_x(l)\hat{S}^{(2)}_x(l)\rangle_{\hat{\rho}_{\text{ST}}}$, we arrive at
\begin{equation}
\label{eq:correlationSxSx_TMST}
\langle\hat{S}^{(1)}_x(l)\hat{S}^{(2)}_x(l)\rangle_{\hat{\rho}_{\text{ST}}} = \frac{2}{\pi}\sum_{m,n=-\infty}^{\infty} X_{n,m},
\end{equation}
where
\begin{equation}
\begin{split}\label{eq:Xnm}
    X_{n,m} \!\!= \!\frac{l\, \Gamma(l,r,T)}{\nu(T)}\!\sqrt{\frac{\pi}{\gamma_2}}\! \int_0^1 \!\!\!\!\dd{z} e^{\frac{-\gamma_1 l^2}{4}(z+2n+2m)\left(z+2n+2m+2\right)} \\
    \hspace{-0.5em}\times\!\! \left[\erf\!\left(\tfrac{l}{2}\sqrt{\gamma_2}(z+2n-2m)\right)\!+ \!\erf\!\left(\tfrac{l}{2}\sqrt{\gamma_2}(z-2n+2m)\right)\right],
\end{split}
\end{equation}
and
\begin{multline}\label{eq:Gamma}
\Gamma(l,r,T) = \exp\left\{-2l^2\nu\!\left(\tfrac{T}{2}\right)e^{-2r}\right\}\\+ \exp\left\{-l^2\left[2\nu\!\left(\tfrac{T}{2}\right)e^{2r} -\tfrac{1}{\nu(T)}\sinh(2r)\right]\right\}\,. 
\end{multline}
Equations \eqref{eq:Znm} and \eqref{eq:Xnm} reduce to the expressions in previous literature~\cite{Main_paper} in the limit $T \rightarrow 0$. The full derivation of these equations have been included in Appx.~\ref{Appendix:Correlation_funcs_TMST}.



\subsection{Evaluating $\langle\hat{B}\rangle$ for the TMS thermal state}

\begin{figure*}
    \centering
    \includegraphics[scale=.6]{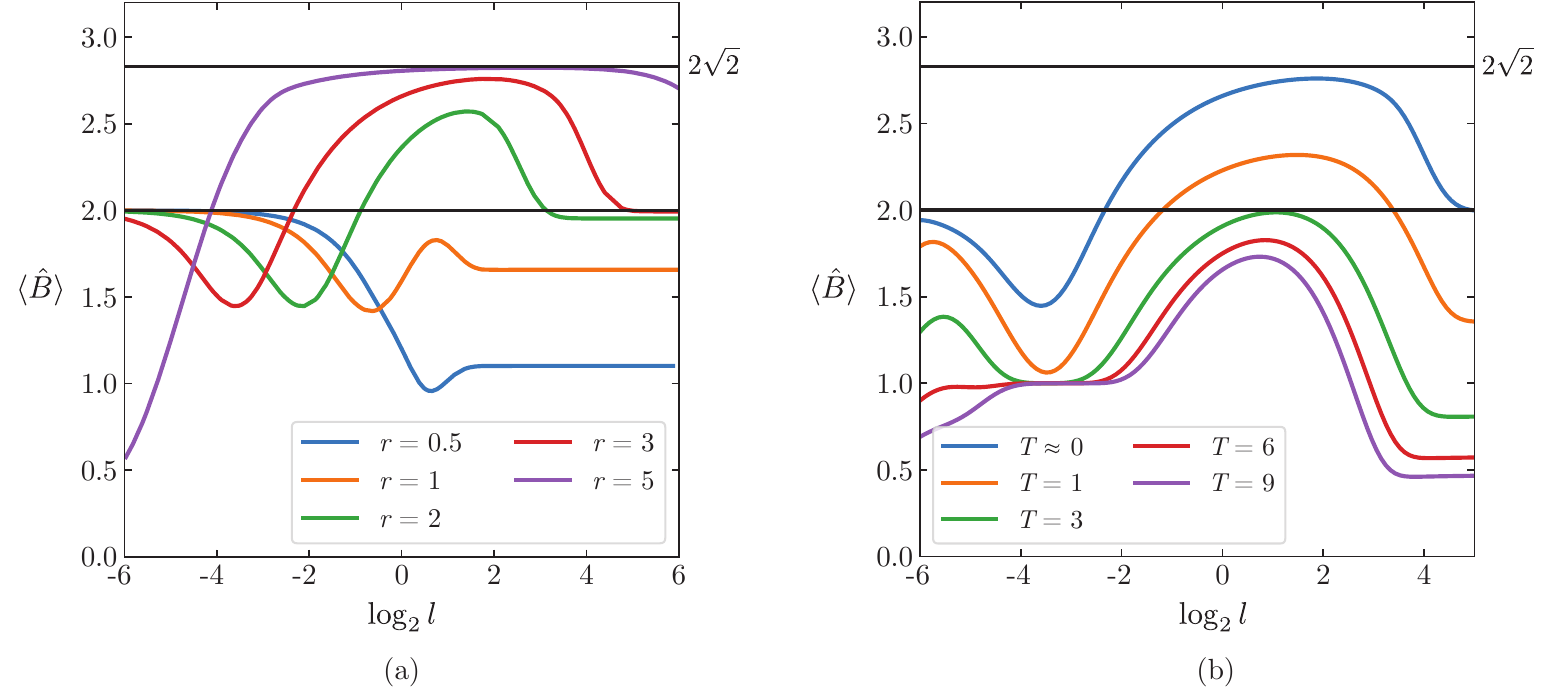}
    \caption{\justifying The expectation value of the Bell operator $\hat{B}$, constructed from the binned pseudospin operators given in Eqs.~\eqref{eq:Sz}--\eqref{eq:Sy}, is plotted for a TMS thermal state as a function of the logarithm of the bin-size $l$ for (a) various squeezing parameters at $T = 10^{-6}$ ($T=0$ is numerically infeasible) and (b)  various temperature values at $r=3$. The former closely resembles the plot in \cite{Main_paper} for the TMS vacuum state. The oscillator frequencies $\omega_A, \omega_B$ are set to 1, and the squeezing angle $\phi$ is set to 0. The line at $\langle\hat{B}\rangle=2$ denotes the boundary between the region of CHSH inequality violation ($\langle\hat{B}\rangle>2$) and non-violation ($\langle\hat{B}\rangle\leq2$) while the line at $\langle\hat{B}\rangle=2\sqrt{2}$ represents Tsirelson's bound.}
    \label{fig:B_l}
\end{figure*}

In this section, we carry out the above integration numerically to examine the dependence of the Bell operator expectation value $\langle\hat{B}\rangle_{\hat{\rho}_{\text{ST}}}$ on the squeezing parameter $r$, temperature $T$, and bin size $l$. 

First, let us revisit the results~\cite{Main_paper} but compute now using the phase space representation for our calculations. In Fig.~\ref{fig:B_l}a, we use the Wigner-Weyl representation of the pseudospin operators to compute the correlation functions in the limit $T \to 0$ to reproduce the results of \cite{Main_paper} for $\langle\hat{B}\rangle$ as a function of $l$ for various values of $r$ in a TMS vacuum state. Like in~\cite{Main_paper} we see that Bell inequality is violated in a finite range of $l$ for sufficiently large $r$. Also, as in~\cite{Main_paper}, we notice that there exists an optimal choice of $l$ that maximizes $\langle\hat B\rangle$ for any given $r$. For the TMS vacuum state, the squeezing cutoff below which we cannot violate Bell inequalities with any choice of $l$ for our choice of pseudospin operators is approximately $r \approx 1.12$. As $r \rightarrow \infty$, $\langle \hat{B} \rangle$ approaches the Tsirelson bound $2\sqrt{2}$ for all $l$, consistent with the fact that in this limit the modes become maximally entangled, approaching the state in the original formulation of the EPR paradox.

After this review, we now study the dependence on temperature of the TMS thermal state. Figure \ref{fig:B_l}b   shows $\langle\hat{B}\rangle_{\hat{\rho}_{\text{ST}}}$ as a function of $l$ for various temperatures at $r = 3$. The maximal BIV decreases rapidly with increasing $T$, showing that thermal noise strongly impedes the observation of BIV. The limit of $\langle\hat{B}\rangle_{\hat{\rho}_{\text{ST}}}$ as $l \to \infty$ also decrease with increasing $T$. This limit is discussed in detail in Appx.~\ref{appx:l_limit}.

\begin{figure}
    \centering
    \includegraphics[scale=.55]{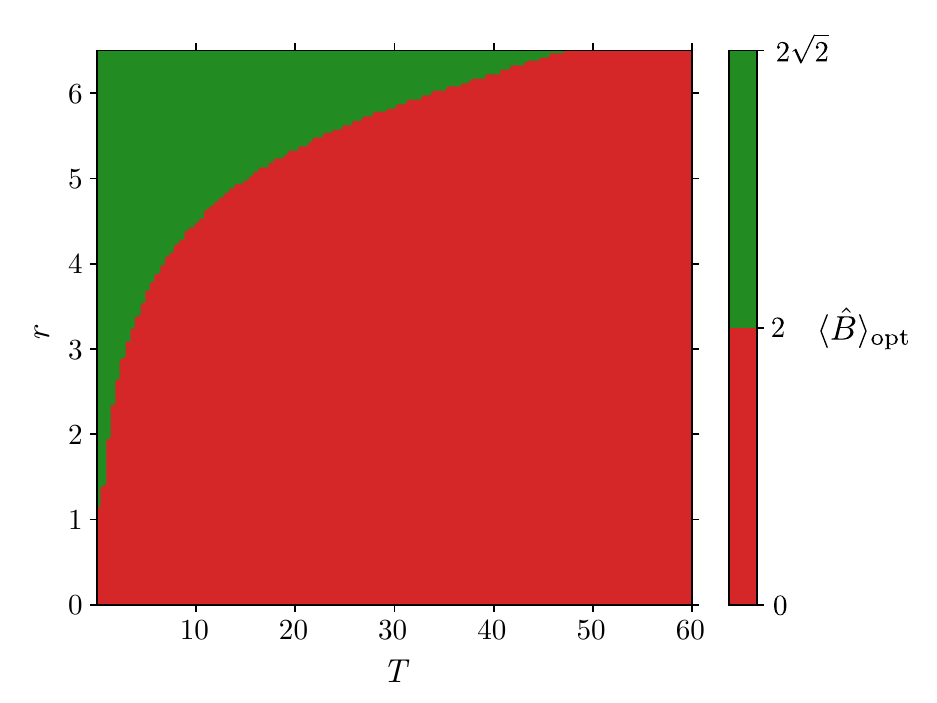}
    \caption{\justifying Regions of CHSH inequality violation for $\langle\hat{B}\rangle$ optimized over bin size $l$ are plotted as a function of the squeezing parameter $r$ and temperature $T$ for a TMS thermal state. The green region indicates violation of the CHSH inequality while the red region indicates non-violation. The free parameters are equivalent to those in Fig.~\ref{fig:B_l}. The violation regions are presented in a binary fashion due to the high computational cost involved in evaluating $\langle\hat{B}\rangle_\text{opt}$, however, curves for fixed values of $r$ and $T$ are shown in Fig.~\ref{fig:B_opt_vary_r_T}.}
    \label{fig:binary_violation_plot}
\end{figure}

\begin{figure*}
    \centering
    \includegraphics[scale=.6]{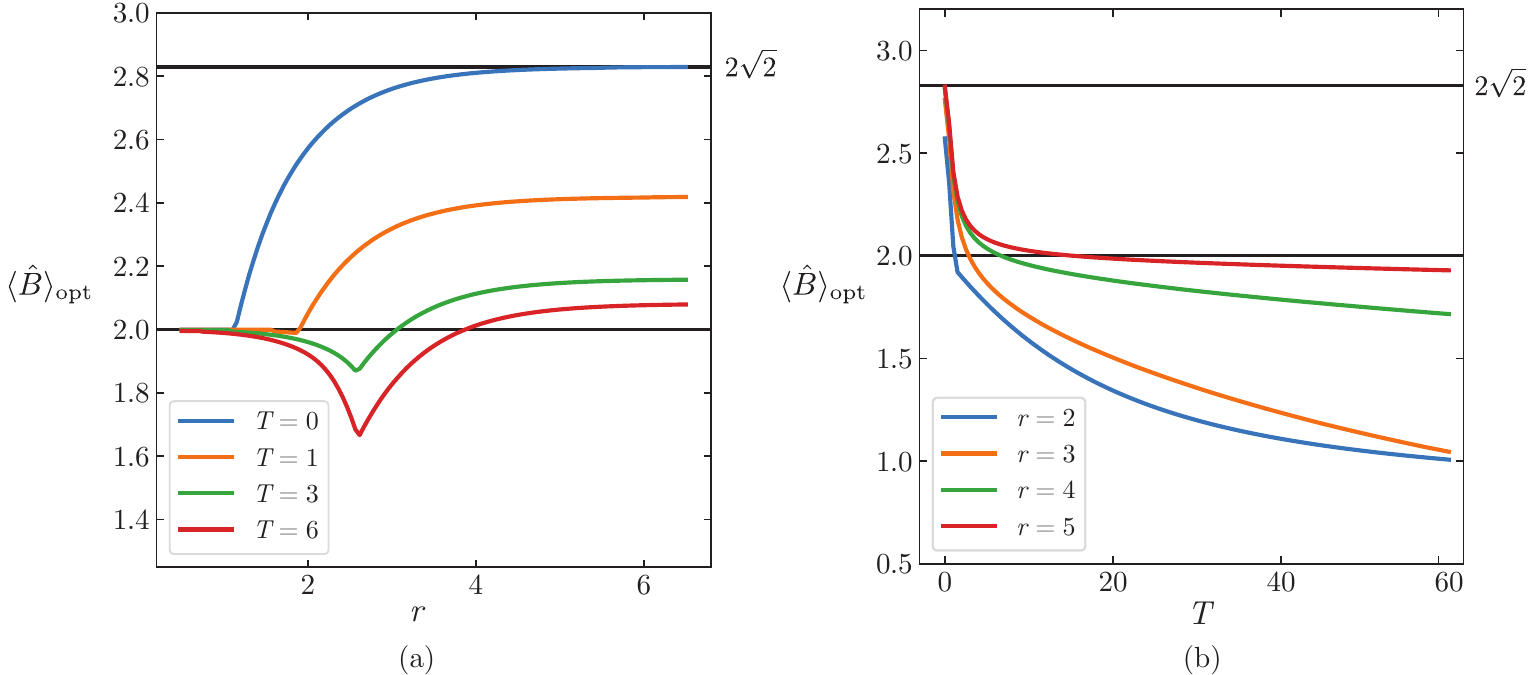}
    \caption{\justifying  $\langle\hat{B}\rangle$, optimized over bin size $l$, is plotted as a function of (a) squeezing parameter $r$ for various fixed values of temperature $T$ and (b) $T$ for various fixed values of $r$. These plots can be seen as fixed-$r$ and fixed-$T$ curves from the data in Fig.~\ref{fig:binary_violation_plot}.}
    \label{fig:B_opt_vary_r_T}
\end{figure*}

Figure~\ref{fig:binary_violation_plot} shows the violation of CHSH inequality (green), or lack thereof (red), as a function of squeezing and temperature for the optimal bin size at each data point. We will denote the use of the optimal $l$ value by $\langle \hat{B} \rangle_\text{opt}$. This value is the maximum of $\langle \hat{B} \rangle_{\hat{\rho}_{\text{ST}}}$ over all possible values of $l$ for a particular choice of $r$ and $T$. Figure~\ref{fig:binary_violation_plot} illustrates that the amount of squeezing necessary to violate the CHSH inequality increases significantly with respect to temperature, a result relevant to experimental tests involving thermal noise. The explicit behaviour of the optimized $\langle \hat{B} \rangle_{\hat{\rho}_{\text{ST}}}$ for fixed values of $r$ and $T$ has been shown in Fig.~\ref{fig:B_opt_vary_r_T}. We observe that while $\langle \hat{B} \rangle_\text{opt}$ varies monotonically with temperature for any fixed value of $r$, it does not show the same monotonic behavior with respect to squeezing at particular values of $T$, except when $T=0$.   

\section{Relationship between entanglement and CHSH violation in the TMS thermal state}\label{sec:entanglement}

A pure bipartite state can violate Bell inequalities if and only if it is entangled \cite{gisin1991bell}. Mixed states, on the other hand, can be entangled but not violate Bell inequalities \cite{Werner}. For this reason, it is particularly important to analyze separately the amount of entanglement and the ability to violate the CHSH inequality of the TMS thermal state: these two features are not necessarily linked given that the TMS thermal state is not pure.
In this section, we examine how $\langle\hat{B}\rangle_{\text{opt}}$ varies along contours of equal entanglement in the TMS thermal state as we vary the squeezing parameter $r$ and the temperature $T$.
\begin{figure}[h]
    \includegraphics[scale=.9]{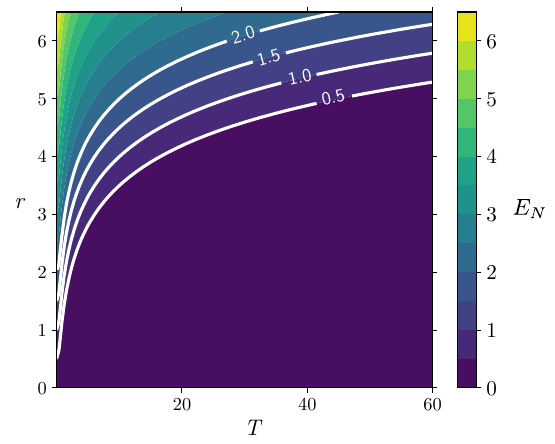}
    \caption{\justifying The logarithmic negativity $E_N$ for a TMS thermal state is plotted as a function of squeezing parameter $r$ and temperature $T$, where the free parameters are equal to those in 
    Fig.~\ref{fig:B_l}. The white lines indicate contours of equal $E_N$.}
    \label{fig:logN_r_T}
\end{figure}

We quantify entanglement for a TMS thermal state with logarithmic negativity \cite{Log_negativity, Plenio}, a faithful entanglement monotone for the TMS thermal state\footnote{Having a non-zero logarithmic negativity is, in general, sufficient but not necessary to have entanglement \cite{Horodecki}. However for certain cases, including the TMS thermal state, logarithmic negativity is actually a faithful entanglement monotone and only vanishes for separable states~\cite{Peres_Horo_Gaussian, Peres_Horo_Gaussian2}.}. The calculation of the logarithmic negativity for the TMS thermal state can be found in Appx.~\ref{appx:cov_and_Wig}. Figure~\ref{fig:logN_r_T} shows the logarithmic negativity of the TMS thermal state as a function of $r$ and $T$. Unsurprisingly, the effect of the thermal noise is to decrease entanglement for fixed $r$~\cite{Isar_entanglement_thermal_squeeze1, Isar_entanglement_thermal_squeeze2}. 

\begin{figure*}[t]
    \centering
    \includegraphics[scale=1]{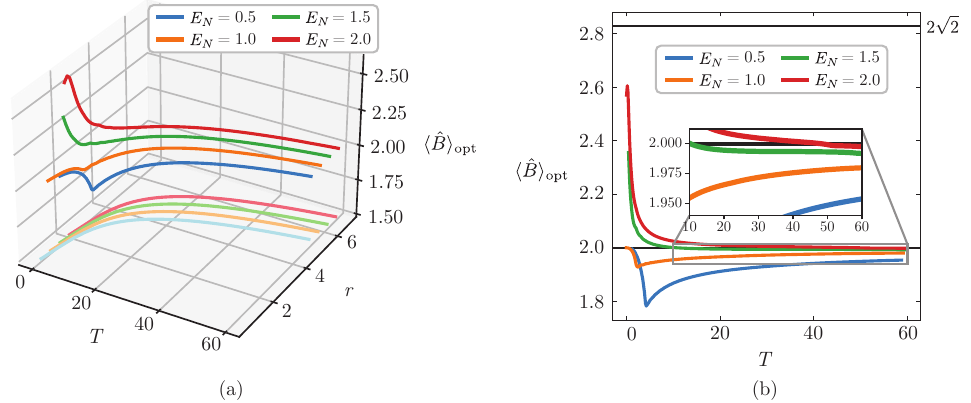}
    \caption{\justifying (a) $\langle\hat{B}\rangle$, optimized over bin size $l$, is plotted as a function of both the squeezing parameter $r$ and temperature $T$, while maintaining constant logarithmic negativity $E_N$. The contours of constant $E_N$ are projected in lighter shades onto the $r-T$ plane, similar to Fig.~\ref{fig:logN_r_T}. (b) A projection of the curves from (a) onto the $\langle\hat{B}\rangle_{\text{opt}}-T$ plane provides a clearer view of the temperature dependence for different values of $E_N$.} 
    \label{fig:entanglement_contours}
\end{figure*}

Figure~\ref{fig:entanglement_contours}(a) shows $\langle\hat{B}\rangle_\text{opt}$ along the equal entanglement contours shown in Fig.~\ref{fig:logN_r_T}. Additionally, we see in Figures~\ref{fig:entanglement_contours}(a)  and~\ref{fig:entanglement_contours}(b) that the curves of constant entanglement are not curves of constant values of $\langle \hat B\rangle_{\text{opt}}$. This should not be surprising, since we know that the violation of the CHSH inequality is not a monotonic function of entanglement  (see e.g., \cite{Gour_position_rep}). 

Figure \ref{fig:entanglement_contours}(b) shows these curves projected onto the $\langle\hat{B}\rangle_{\text{opt}}-T$ plane. Let us draw attention to the comparison of states with the same entanglement but different values of $r$ and $T$. In particular, for the states with $E_N=1.5$ and $E_N=2$, we see that lower values of $r$ and $T$ can violate CHSH whereas larger values of $r$ and $T$ cannot, despite having the same entanglement. 
Notice as well that the curves that do not exhibit violations of CHSH at lower values of $r$ and $T$ (i.e., with $\langle\hat{B}\rangle_{\text{opt}} < 2$) gradually increase as $r$ and $T$ grow until they reach a maximum, but this maximum is never above $\langle\hat{B}\rangle_{\text{opt}}=2$.

The sharp variation of $\langle\hat{B}\rangle_{\text{opt}}$ for lower values of $r$ and $T$ is due to the fact that the optimal choice of $l$ that maximizes the CHSH violation is actually very small, and at some point when $r$ and $T$ are increased the optimal value discontinuously jumps to a larger value. This is illustrated for the TMS vacuum in Fig.~\ref{fig:B_l}a, where we see the maximum of $\langle\hat{B}\rangle$ for $r=1$ happens already at the asymptote of very low $l$ whereas for higher squeezing the maximum happens at finite values of $l$ (in the `bump' region).


Several inferences can be drawn from Fig.~\ref{fig:entanglement_contours}. First, $\langle\hat{B}\rangle_{\text{opt}}$ is not a monotonic function of entanglement for the choice of pseudospin operators \eqref{eq:Sz}--\eqref{eq:Sy} and the TMS thermal state. Second, for constant entanglement, the expectation value of the Bell operator remains close to the boundary of the violation region $\langle\hat{B}\rangle_{\text{opt}}=2$ for most values of $r$ and $T$. Third, the curves outside the violation region initially increase with increasing $r$ and $T$ at constant entanglement, implying that squeezing has a greater effect than temperature. Fourth, as we move further along the equal entanglement contours to higher $r$ and $T$ values, $\langle\hat{B}\rangle_{\text{opt}}$ falls below 2 deeper into the classical region, suggesting that the effect of temperature outweighs that of squeezing in this region. This is further illustrated in Fig.~\ref{fig:B_opt_vary_r_T}, where we showed that the behavior of the violations of CHSH inequalities is not strictly monotonic with the amount of squeezing in the state at constant temperature, and that a large amount of squeezing is necessary to compensate for the reduction of $\langle \hat{B}\rangle_\text{opt}$ caused by increasing temperatures.

Naively, one might expect that $\langle\hat{B}\rangle_{\text{opt}}$ would remain constant as long as $T$ and $r$ vary so as to keep the amount of entanglement constant. However, this is not the case, and there are two potential explanations for this behavior. First, even after optimizing over the parameter $l$, the optimal choice of pseudospin operators for the TMS thermal state remains unclear---a point we will explore further in the next section. Second, the TMS thermal state is not pure and may exhibit behavior similar to the Werner state as it becomes more mixed: it may be losing its ability to violate the CHSH inequalities even if it is entangled.

\section{Optimality and choice of pseudospin operators}
\label{sec:operator_choice}

Finding optimal pseudospin operators for continuous variable systems is a challenging task due to the infinite-dimensional nature of the Hilbert space of those systems. Nevertheless, \cite{Gour_position_rep} showed that for any choice of triplets of dichotomous  operators $\{\hat{\bm{\sigma}}_i\}$ on the Hilbert space of a continuous variable system which satisfies the $\mathfrak{su}(2)$ algebra, the TMS vacuum state obeys
\begin{equation}\label{eq:pseudospin_Bell}
    \langle\hat{B} \rangle_{\hat{\rho}_{\text{SV}}} \leq 2\sqrt{1+\tanh^2(2r)}\,.
\end{equation}
This bound is saturated for the TMS vaccum for all values of $r$ by the operators presented in \cite{Chen_pseudospin} (see Eq.~\eqref{eq:pseudo_fock}). Therefore, this set of pseudospin operators is optimal for the TMS vacuum state (although not necessarily the only optimal set). Note that the fact that the operators in \cite{Chen_pseudospin} saturate Eq.~\eqref{eq:pseudospin_Bell} implies that (for that choice of pseudospin operators) the CHSH inequality is violated for all $r >0$ and maximally violated in the limit of infinite squeezing $r \to \infty$. However, recall from Sec.~\ref{section:operator_presentation} that these operators are presented in the Fock representation, making it challenging to use them in a phase space formulation of the CHSH inequality violation.

A different set of pseudospin operators (those introduced in~\cite{Gour_position_rep} and presented in Eq.~\eqref{eq:pseudo_pos}) also violate the CHSH inequality for all $r >0$ and yield maximal violation for $r\to\infty$. Conveniently, those operators are given in the position representation but they do not saturate the above bound for all values of $r$. When it comes to finding pseudospin operators expressed in the position representation, these are the closest to saturating the inequality~\eqref{eq:pseudospin_Bell} for the TMS vacuum state that have been explored so far. 

\begin{figure}
    \centering
    \includegraphics[scale=0.9]{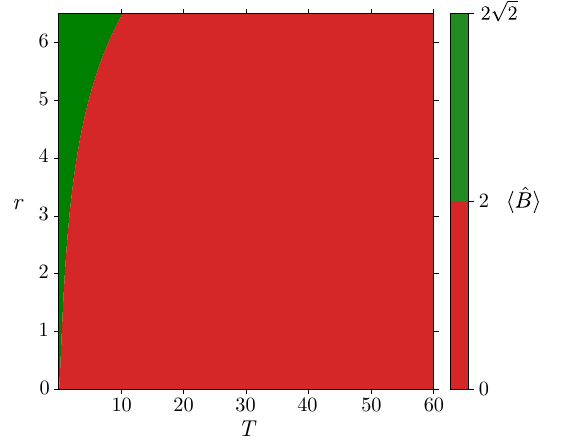}
    \caption{\justifying Regions of violation (green) and non-violation (red) of the CHSH inequality are plotted as a function of the squeezing parameter $r$ and temperature $T$ for the TMS thermal state evaluated using the unbinned pseudospin operators given in Eq.~\eqref{eq:pseudo_pos}.}
    \label{fig:binary_violation_plot_2}
\end{figure}

There is yet another proposal for pseudospin operators that is easy to express in terms of the position basis, namely the binned operators introduced in~\cite{Main_paper2} and used throughout this paper (Eqs.~\eqref{eq:Sz}-\eqref{eq:Sy}). For the TMS vacuum state, this choice turns out to be outperformed by the (unbinned) pseudospin operators \eqref{eq:pseudo_pos}:  there is no choice of binning size $l$ that will violate the CHSH inequality for $r=0$ to $r \approx 1.12$, as can be seen in Fig.~\ref{fig:binary_violation_plot}. However, for the TMS thermal state, the binned operator choice in~\cite{Main_paper2} actually outperforms the unbinned choice in \cite{Gour_position_rep} for large enough temperatures. One can see this by comparing Fig.~\ref{fig:binary_violation_plot_2}, which shows $\langle\hat{B}\rangle_{\hat{\rho}_{\text{ST}}}$ evaluated using unbinned pseudospin operators \eqref{eq:pseudo_pos}, with Fig.~\ref{fig:binary_violation_plot}. This highlights that choices of operators that may be optimal for the TMS vacuum may not be so for thermal states. The computation of $\langle\hat{B}\rangle_{\hat{\rho}_{\text{ST}}}$ using unbinned pseudospin operators  (Fig.~\ref{fig:binary_violation_plot_2}) has been included in Appx.~\ref{appendix:unbinned_Bell_op}.




A larger discussion should include finding the optimal operators for each value of temperature. Nothing guarantees that any of the previous literature choices are optimal for an arbitrary temperature. 

In trying to find the optimal operators for each temperature, one can quickly check that other pseudospin operators choices sharing the same $\hat S_z$, as in~\cite{Main_paper2}, but rotating the $\hat S_x$ and $\hat S_y$ in the Bloch sphere will not yield improvements. Namely, if we choose
\begin{align}
    &\hat{\tilde{S}}^{(i)}_x = \cos\theta \hat{S}^{(i)}_x - \sin\theta \hat{S}^{(i)}_y,\\
    &\hat{\tilde{S}}^{(i)}_y = \sin\theta \hat{S}^{(i)}_x + \cos\theta \hat{S}^{(i)}_y,
\end{align}
one can see that
\begin{equation}
    \langle \hat{\tilde{S}}^{(1)}_x \hat{\tilde{S}}^{(2)}_x \rangle_{\hat{\rho}}^2 = (\cos 2\theta)^2 \langle \hat{S}^{(1)}_x \hat{S}^{(2)}_x \rangle_{\hat{\rho}}^2,
\end{equation}
which is maximized for $\theta = 0$ and thereby, so is $\langle \hat{B}\rangle$ through Eq.~\eqref{eq:Bell_op}.

%
%
%
%

Unlike in the case of spin 1/2 systems---for which the entire space of sets of Pauli operators can be generated by Bloch sphere rotations---for continuous variables there are infinitely many choices of pseudospin operators that are not equivalent under Bloch sphere rotations. This is not unexpected since a set of three operators spans only a small subspace of our total space of operators in the infinite-dimensional Hilbert space. 

All considered, in this work, we have shown that the binned operators are closer to optimality than any other previously proposed choice of operators presented in the literature thus far. The question remains open as to what would be the operators that optimize the violation of CHSH inequalities for two-mode squeezed thermal states. 
\vspace{4mm}


\section{Conclusion}\label{sec:conclusion}

We presented a formalism to evaluate the violation of local realism for mixed Gaussian states by expressing the binned pseudospin operators \cite{Main_paper2} in the Wigner-Weyl representation. This formalism allowed us to compute the violation of the CHSH inequality for a two-mode squeezed (TMS) thermal state. We showed that as the temperature of a TMS thermal state increases, substantially greater squeezing is required to violate the CHSH inequality. Notably, this required squeezing surpasses the amount needed to maintain a constant level of entanglement in the state as the temperature rises. This result is relevant for tests of local realism involving thermal noise. 

To study the relationship between violations of the CHSH inequality and entanglement, we analyzed the expectation value of the Bell operator at fixed entanglement for varying values of $r$ and $T$. We observed that the magnitude of the violation is not monotonic along curves of constant entanglement for the choice of pseudospin operators that we considered. Finally, we showed that the hierarchy of operators that are optimal for the TMS vacuum state is not respected once temperature is considered. Namely, the binned pseudospin operators, which were suboptimal for the TMS vacuum, show stronger violations of the CHSH inequality for the TMS thermal state.

A future direction involves finding optimal pseudospin operators for each temperature for the TMS thermal state. This endeavor will be valuable for discerning to what extent the non-monotonic behavior of the CHSH violations with entanglement comes from the choice of operators, the mixedness of the TMS thermal state, or both. In this regard, one can also attempt to determine if there is an optimal choice of binned pseudospin operators (following the philosophy of Larsson's~\cite{Main_paper2}) such that in the low-temperature regime close to $T\rightarrow0$, they saturate the inequality~\eqref{eq:pseudospin_Bell}. This is promising since, as we showed, the Larsson binned operators outperform other choices for large enough temperatures after optimizing over bin size. This approach is currently under investigation.

\bigskip

\begin{center}
    \textbf{ACKNOWLEDGEMENT}
\end{center}
The authors thank T. Rick Perche for helpful discussions. GS thanks the PSI Masters program at Perimeter Institute for facilitating this research. E. M-M. is funded by the NSERC Discovery program as well as his Ontario Early Researcher Award. Research at Perimeter Institute is supported in part by the Government of Canada through the Department of Innovation, Science and Industry Canada and by the Province of Ontario through the Ministry of Colleges and Universities. Perimeter Institute and the University of Waterloo are
situated on the Haldimand Tract, land that was promised to the Haudenosaunee of the Six Nations of the Grand River, and is within the territory of the Neutral, Anishinaabe, and Haudenosaunee people.

\appendix

\section{Pseudospin operators in the Fock representation}
\label{appx:pseudospin_ops}
In this appendix, we present the pseudospin operators referenced in Sec.~\ref{subsec:binned_pos_op} but not explicitly shown there. These operators are provided in literature in their Fock representation.

The operators that have been proven optimal for the TMS vacuum state were first presented in \cite{Chen_pseudospin} in the Fock representation as
\begin{subequations}
\label{eq:pseudo_fock}
\begin{align}
    \hat{s}_z &= \sum_{n=0}^{\infty} \left\{\ket{2n+1}\!\bra{2n+1}-\ket{2n}\!\bra{2n}\right\} \\ 
    \hat{s}_x &= \sum_{n=0}^{\infty} \left\{\ket{2n+1}\!\bra{2n}+\ket{2n}\!\bra{2n+1}\right\} \\
    \hat{s}_y &= -\ii\sum_{n=0}^{\infty} \left\{\ket{2n+1}\!\bra{2n}-\ket{2n}\!\bra{2n+1}\right\}\,.
\end{align}
\end{subequations}
They obey the $\mathfrak{su}(2)$ algebra, making them a bosonic analog to the Pauli operators for spin-1/2 systems. If $n=0$, we recover the original Pauli operators. For optimal measurement angles in $\langle \hat{B} \rangle$ (see Eq.~\eqref{eq:n_vector}) and for the TMS vacuum state $\hat{\rho}_{\text{SV}}$, we get
\begin{align}\label{eq:Bell_Chen}
    \langle\hat{B}\rangle_{\hat{\rho}_{\text{SV}}} = 2\sqrt{1+(F(r))^2} \,,
\end{align}
where 
\begin{align}
\label{eq:Fr_Chen}
    F(r) = \langle\hat{s}^{(1)}_x\otimes\hat{s}^{(2)}_x\rangle_{\hat{\rho}_{\text{SV}}}
    = \tanh(2r)\,.
\end{align}
The above operators saturate the inequality in Eq.~\eqref{eq:pseudospin_Bell} and thus maximally violate the CHSH inequality for the TMS vacuum state for all values of $r$. 

In \cite{Gour_position_rep}, another set of pseudospin operators was introduced, expressed in the position space representation. We presented these operators in Eq.~\eqref{eq:pseudo_pos}.
Using these operators, $F(r)$ for optimized measurement angles in $\langle \hat{B} \rangle$ comes out to be 
\begin{equation}
    F(r) = \langle\hat{\Pi}^{(1)}_x\otimes\hat{\Pi}^{(2)}_x\rangle_{\hat{\rho}_{\text{SV}}}
    = \frac{2}{\pi}\arctan(\sinh(2r))\,.
\label{eq:Bell_Gour}
\end{equation}
Note that while Eq.~\eqref{eq:Bell_Gour} matches Eq.~\eqref{eq:Fr_Chen} for $r=0$ and for $r \to \infty$, at all intermediary values, the $F(r)$ obtained with the operators expressed in the position representation is strictly less than the (optimal) operators expressed in the Fock representation ($\frac{2}{\pi}\arctan(\sinh(2r))< \tanh(2r)$ for $r\in (0,\infty)$). The position-space representation of the optimal operators for the TMS vacuum state \eqref{eq:pseudo_fock} has not yet been presented in the literature to the authors' knowledge.

The binned pseudospin operators in the Fock representation were introduced by Larsson in \cite{Larsson_fock_basis_grouping} and are constructed by grouping $d$ Fock states in Eq.~\eqref{eq:pseudo_fock} together. These are given by
\begin{subequations} \label{eq:pseudo_fock_group}
\begin{align}
    \hat{s}_{z,d} &= \sum_{n=0}^{\infty} (-1)^n \sum_{k=0}^{d-1} \ket{dn+k}\!\bra{dn+k} \label{eq:pseudo_fock_group_z} \\ 
    \hat{s}_{x,d} &= 
    \begin{aligned}[t]
        &\sum_{n=0}^{\infty} \sum_{k=0}^{d-1} \big[\ket{2dn+k}\!\bra{2dn+k+d} \\
        &\phantom{={}}+\ket{2dn+k+d}\!\bra{2dn+k}\big]
    \end{aligned} \label{eq:pseudo_fock_group_x} \\
    \hat{s}_{y,d} &= 
    \begin{aligned}[t]
        &-\ii\sum_{n=0}^{\infty} \sum_{k=0}^{d-1} \big[\ket{2dn+k}\!\bra{2dn+k+d} \\
        &\phantom{={}}-\ket{2dn+k+d}\!\bra{2dn+k}\big]\,.
    \end{aligned} \label{eq:pseudo_fock_group_y}
\end{align}
\end{subequations}
With these operators, $\langle \hat{B} \rangle_{\hat{\rho}_{\text{SV}}}$ depends on both squeezing parameter $r$ as well as the group size $d$. For the optimal set of measurement angles, $\langle \hat{B} \rangle_{\hat{\rho}_{\text{SV}}}$ is given by \cite{pseudospin_degerency}
\begin{equation}
\label{eq:Bell_pseudo_fock_group}
\langle\hat{B}\rangle_{\hat{\rho}_{\text{SV}}} = \frac{2\sqrt{\tanh^{4d}(r) + 6\tanh^{2d}(r)+1}}{1+\tanh^{2d}(r)}\,.
\end{equation}
For $d=1$, Eqs.~\eqref{eq:pseudo_fock_group} and \eqref{eq:Bell_pseudo_fock_group} coincide with Eqs.~\eqref{eq:pseudo_fock} and \eqref{eq:Bell_Chen} respectively. However, notice that this is unlike the (position space) binned pseudospin operators (Eqs.~\eqref{eq:Sz}--\eqref{eq:Sy}, which form the basis of our work) which do not coincide with the unbinned pseudospin operators \eqref{eq:pseudo_pos} for any bin-size $l$.  

\section{Correlation functions for binned pseudospin operators in Wigner-Weyl representation}
\label{appx:correlation_funcs}
In Sec.~\ref{subsec:correlation_funcs_Wigner}, the correlation functions $ 
\langle\hat{S}^{(1)}_z(l)\hat{S}^{(2)}_z(l)\rangle_{\hat{\rho}} $, $\langle\hat{S}^{(1)}_+(l)\hat{S}^{(2)}_+(l)\rangle_{\hat{\rho}} $ and $ \langle\hat{S}^{(1)}_+(l)\hat{S}^{(2)}_-(l)\rangle_{\hat{\rho}} $ were given in terms of the Wigner function $W_{\hat{\rho}}$ in Eqs.~\eqref{eq:correlationSzSz}, \eqref{eq:correlationS+S+}
and \eqref{eq:correlationS+S-} respectively. In this appendix, we will derive these expressions as well as the expressions for the remaining correlation functions. 

To calculate $ \langle\hat{S}^{(1)}_z(l)\hat{S}^{(2)}_z(l)\rangle_{\hat{\rho}} $, we must first derive the Wigner-Weyl (phase-space) representation of $\hat{S}_z^{(i)}(l)$ (where $i = 1,2$) which was given in Eq.~\eqref{eq:Sz_Wig_trans}. Eq.~\eqref{eq:Wig_trans} provides the general form of the Wigner-Weyl representation of an operator, which in this case yields
\begin{align}
    &\hspace{-2em}\phantom{={}}S_z^{(i)}(l,q_i,p_i)\! = \!\!\int_{-\infty}^{\infty} \!\!\!\!\!\!\!\dd{x} e^{-\ii p_i x}\! \mel{q_i+\tfrac{1}{2}x}{\hat{S}^{(1)}_z(l)}{q_i - \tfrac{1}{2}x}\\
    &= \int_{-\infty}^{\infty}\!\!\!\! \dd x e^{-\ii p_1 x} \sum_{n=-\infty}^{\infty} (-1)^n \int_{nl}\limits^{(n+1)l} \dd{q}\delta\left(q_i+\tfrac{1}{2}x-q\right)\nonumber\\\label{eq:step1}
    &\hspace{45mm}\times \delta\left(q_i-\tfrac{1}{2}x-q\right) . 
\end{align}
Using $\delta(f(x)) = \frac{\delta(x-x_0)}{|f'(x_0)|}$, we get
\begin{align}\label{eq:Sz1}
 S_z^{(i)}(l,q_i,p_i) &=4 \sum_{n=-\infty}^{\infty} (-1)^n \int_{nl}\limits^{(n+1)l} \dd{q} \int_{-\infty}^{\infty} \dd{x} e^{-\ii p_i x} \nonumber\\
 &\quad\times\delta(x-2q+2q_i)\delta(x-2q_i+2q) \nonumber\\
 &\hspace{-4em}\begin{aligned}[t]
    = \sum_{n=-\infty}^{\infty} (-1)^n \int_{nl}\limits^{(n+1)l} \dd{q} e^{-2\ii p_i(q-q_i)} \delta(q-q_i) \,.
\end{aligned}
\end{align}
Using \eqref{eq:expectation_value_Wigner}, we can calculate the correlation function $\langle\hat{S}^{(1)}_z(l)\hat{S}^{(2)}_z(l)\rangle_{\hat{\rho}}$ for a state $\hat{\rho}$ with Wigner representation $W_{\hat{\rho}}$:
\begin{align}\label{eq:correlationSzSz1}
\left\langle\hat{S}^{(1)}_z(l)\hat{S}^{(2)}_z(l)\right\rangle_{\hat{\rho}} &= \int_{-\infty}^{\infty}\!\!\! \dd{q_1} \dd{q_2} \dd{p_1}\dd{p_2} W_{\hat{\rho}}(q_1,q_2,p_1,p_2) \nonumber\\
 &\quad\times S_z^{(1)}(l,q_1,p_1)S_z^{(2)}(l,q_2,p_2) \\
&\hspace{-7em}=\!\!\!\sum_{m,n=-\infty}^{\infty}\!\!\!(-1)^{m+n} \int_{-\infty}^{\infty}\!\!\!\dd{q_1}\dd{q_2} \dd{p_1}\dd{p_2} W_{\hat{\rho}}(q_1,q_2,p_1,p_2) \nonumber \\ 
&\hspace{-7.5em}\phantom{={}}\!\!\! \times\!\!\!\!\!\int_{nl}\limits^{(n+1)l} \!\!\!\!\!\!\dd{q} \int_{ml}\limits^{(m+1)l} \!\!\!\!\!\!\dd{q'} e^{-2\ii(p_1(q-q_1) + p_2(q'-q_2))}\delta(q-q_1)\delta(q'-q_2) \,.\vphantom{\Bigg\rbrace} 
\end{align}
Integrating over $q_1$ and $q_2$ with the delta functions reduces the exponent of the exponential to zero and replaces the argument of the Wigner function with $q$, $q'$ in place of $q_1$, $q_2$:
\begin{align}\label{eq:correlationSzSz2}
\hspace{-1em}\left\langle\hat{S}^{(1)}_z(l)\hat{S}^{(2)}_z(l)\right\rangle_{\hat{\rho}} &=\!\!\!\!\sum_{m,n=-\infty}^{\infty}(-1)^{m+n} \int_{-\infty}^{\infty}\dd{p_1}\dd{p_2}\nonumber\\
&\hspace{-1em}\quad\times\!\!\!\!\! \int_{nl}\limits^{(n+1)l} \!\!\!\!\!\dd{q} \int_{ml}\limits^{(m+1)l} \!\!\!\!\!\dd{q'} W_{\hat{\rho}}(q,q',p_1,p_2)\,.
\end{align}
Relabeling the integration variables $q\to q_1$ and $q'\to q_2$, we arrive at Eq.~\eqref{eq:correlationSzSz}.
%

Let us now turn to $ \langle\hat{S}^{(1)}_+(l)\hat{S}^{(2)}_+(l)\rangle_{\hat{\rho}} $. The Wigner-Weyl representation of $\hat{S}^{(i)}_+(l)$ is given by 
\begin{align}
    S^{(i)}_+(l,q_i,p_i) &= \int_{-\infty}^{\infty}\!\! \dd{x} e^{-\ii p_i x} \mel{q_i+\tfrac{1}{2}x}{\hat{S}^{(1)}_+(l)}{q_i - \tfrac{1}{2}x}  \nonumber \\
    &=  \sum_{n=-\infty}^{\infty}  \int_{2nl}\limits^{(2n+1)l}\!\!\!\! \dd{q} \int_{-\infty}^{\infty}\!\! \dd x e^{-\ii p_i x}\nonumber\\
&\quad\times\delta\left(q_i+\tfrac{1}{2}x-q\right)\delta\left(q_i-\tfrac{1}{2}x-q-l\right) \nonumber\\
    \label{eq:S+2}
    & =\!\!\!\!  \sum_{n=-\infty}^{\infty} \!\!\! \int_{2nl}\limits^{(2n+1)l}\!\!\!\! \dd{q} e^{-2\ii p_i (q-q_i)}\delta\left(q_i-q-\tfrac{l}{2}\right)\,.
\end{align}
From here, the correlation function $\langle\hat{S}^{(1)}_+(l)\hat{S}^{(2)}_+(l)\rangle_{\hat{\rho}}$ can be obtained by following the steps outlined in Eqs.~\eqref{eq:correlationSzSz1} through \eqref{eq:correlationSzSz2}. The final expression is
\begin{align}\label{eq:correlationS+S+2}
    \left\langle\hat{S}^{(1)}_+(l)\hat{S}^{(2)}_+(l)\right\rangle_{\hat{\rho}} &=\!\!\! \sum_{m,n=-\infty}^{\infty} \int_{-\infty}^{\infty}\!\!\!\!\dd{p_1}\dd{p_2}\!\!\!\int_{2nl}\limits^{(2n+1)l}\!\!\!\! \dd{q_1} \int_{2ml}\limits^{(2m+1)l} \!\!\!\!\dd{q_2} \nonumber\\
&\hspace{-3.5em}\quad\times e^{\ii(p_1+p_2)l}\, W_{\hat{\rho}}(q_1+\tfrac{l}{2},q_2+\tfrac{l}{2},p_1,p_2)\,,
\end{align}
which matches Eq.~\eqref{eq:correlationS+S+}.

Now, we repeat the same steps for $\langle\hat{S}^{(1)}_+(l)\hat{S}^{(2)}_-(l)\rangle_{\hat{\rho}} $. From Eqs.~\eqref{eq:S-} and \eqref{eq:Wig_trans}, the Wigner representation of $\hat{S}^{(2)}_-(l)$ is found to be
\begin{equation}
    S^{(2)}_-(l,q_2,p_2) =  \!\!\!\!\sum_{m=-\infty}^{\infty}  \int_{(2m+1)l}\limits^{(2m+2)l}\!\!\!\!\dd{q} e^{-2\ii p_2 (q-q_2)}\delta\left(q_2-q+\tfrac{l}{2}\right)\,.
\end{equation}
Again performing a similar calculation as in Eqs.~\eqref{eq:correlationSzSz1} through \eqref{eq:correlationSzSz2}, we arrive at
\begin{align}\label{eq:correlationS+S-2}
    \left\langle\hat{S}^{(1)}_+(l)\hat{S}^{(2)}_-(l)\right\rangle_{\hat{\rho}} &=\!\!\! \sum_{m,n=-\infty}^{\infty} \int_{-\infty}^{\infty}\!\!\!\!\dd{p_1}\dd{p_2} \int_{2nl}\limits^{(2n+1)l} \!\!\!\!\dd{q_1} \int_{(2m+1)l}\limits^{(2m+2)l} \!\!\!\!\dd{q_2}\nonumber\\
&\hspace{-4em}\quad\times e^{\ii(p_1-p_2)l}\, W_{\hat{\rho}}(q_1+\tfrac{l}{2},q_2-\tfrac{l}{2},p_1,p_2)\,.
\end{align}
Thus, from Eqs.~\eqref{eq:correlationS+S+2} and \eqref{eq:correlationS+S-2}, we can obtain the expression for $\langle\hat{S}^{(1)}_x(l)\hat{S}^{(2)}_x(l)\rangle_{\hat{\rho}}$ by substituting in Eq.~\eqref{eq:correlationSxSx}.

Next, let us do the same for $\langle\hat{S}^{(1)}_z(l)\hat{S}^{(2)}_x(l)\rangle_{\hat{\rho}}$. Since $\hat{S}_x(l) = \hat{S}_+(l)+ \hat{S}_-(l)$, we have 
$$\left\langle\hat{S}^{(1)}_z(l)\hat{S}^{(2)}_x(l)\right\rangle_{\hat{\rho}} = \left\langle\hat{S}^{(1)}_z(l)\hat{S}^{(2)}_+(l)\right\rangle_{\hat{\rho}} + \left\langle\hat{S}^{(1)}_z(l)\hat{S}^{(2)}_-(l)\right\rangle_{\hat{\rho}}\,. $$
But since $\hat{S}_-(l) = \hat{S}_+^{\dagger}(l)$, we can write
\begin{equation}\label{eq:correlationSzSx}
    \left\langle\hat{S}^{(1)}_z(l)\hat{S}^{(2)}_x(l)\right\rangle_{\hat{\rho}} = 2\Re\Big\{ \left\langle\hat{S}^{(1)}_z(l)\hat{S}^{(2)}_+(l)\right\rangle_{\hat{\rho}}\Big\}\,.
\end{equation}
We can express $\langle\hat{S}^{(1)}_z(l)\hat{S}^{(2)}_+(l)\rangle_{\hat{\rho}}$ in the Wigner representation as: 
\begin{align}
    \left\langle\hat{S}^{(1)}_z(l)\hat{S}^{(2)}_+(l)\right\rangle_{\hat{\rho}} &= \int_{-\infty}^{\infty}\!\!\!\! \dd{q_1} \dd{q_2} \dd{p_1}\dd{p_2} W_{\hat{\rho}}(q_1,q_2,p_1,p_2)\nonumber\\
    &\quad\times S_z^{(1)}(l,q_1,p_1)\,S_+^{(2)}(l,q_2,p_2) \nonumber
\end{align}
From Eqs.~\eqref{eq:Sz1} and \eqref{eq:S+2}, we can write 
\begin{align}
    &=\!\!\sum_{m,n=-\infty}^{\infty}\!\!(-1)^{n}\!\! \int_{-\infty}^{\infty}\!\!\!\!\dd{q_1}\dd{q_2} \dd{p_1}\dd{p_2} W_{\hat{\rho}}(q_1,q_2,p_1,p_2)\!\!\int_{nl}\limits^{(n+1)l} \!\!\!\!\!\dd{q} \nonumber\\
    &\hspace{-1em}\quad\times\!\!\!\!\int_{2ml}\limits^{(2m+1)l} \!\!\!\!\!\!\dd{q'} e^{-2\ii(p_1(q-q_1) + p_2(q'-q_2))}\delta(q-q_1)\delta(q_2-q'-\tfrac{l}{2}) \,.\vphantom{\Bigg\rbrace}\nonumber\\
\label{eq:correlationSzS+}
   &=\!\!\sum_{m,n=-\infty}^{\infty}\!\!(-1)^{n}\!\! \int_{-\infty}^{\infty}\!\!\!\!\dd{p_1}\dd{p_2} e^{\ii p_2 l}\int_{nl}\limits^{(n+1)l} \!\!\!\!\dd{q_1}\int_{2ml}\limits^{(2m+1)l} \!\!\!\!\dd{q_2} \nonumber\\
    &\quad\times W_{\hat{\rho}}(q_1,q_2+\tfrac{l}{2},p_1,p_2)\,. 
\end{align}
Hence, we get $\langle\hat{S}^{(1)}_z(l)\hat{S}^{(2)}_x(l)\rangle_{\hat{\rho}}$ correlation function by substituting in Eq.~\eqref{eq:correlationSzSx}. 

The derivation for spin-$z$, spin-$y$ two-point correlation functions is also along the same lines. From Eq.~\eqref{eq:Sy}, the $\langle\hat{S}^{(1)}_y(l)\hat{S}^{(2)}_z(l)\rangle_{\hat{\rho}}$ can be written as 
$$\langle\hat{S}^{(1)}_y(l)\hat{S}^{(2)}_z(l)\rangle_{\hat{\rho}} = -\ii\langle\hat{S}^{(1)}_+(l)\hat{S}^{(2)}_z(l)\rangle_{\hat{\rho}} +\ii \langle\hat{S}^{(1)}_-(l)\hat{S}^{(2)}_z(l)\rangle_{\hat{\rho}}\,. $$
Again from $\hat{S}_-(l) = \hat{S}_+^{\dagger}(l)$, we have
\begin{equation}\label{eq:correlationSzSy}
    \langle\hat{S}^{(1)}_y(l)\hat{S}^{(2)}_z(l)\rangle_{\hat{\rho}} = 2\Im\Big\{ \langle\hat{S}^{(1)}_+(l)\hat{S}^{(2)}_z(l)\rangle_{\hat{\rho}}\Big\} \,.
\end{equation}
We already know the correlation function on r.h.s from Eq.~\eqref{eq:correlationSzS+} (we just have to swap the label (1) and (2)). 

Next, we move onto $\langle\hat{S}^{(1)}_x(l)\hat{S}^{(2)}_y(l)\rangle_{\hat{\rho}}$ which can be found using Eqs.~\eqref{eq:Sx} and \eqref{eq:Sy} to be 
\begin{align}\label{eq:correlationSxSy}
   \langle\hat{S}^{(1)}_x(l)\hat{S}^{(2)}_y(l)\rangle_{\hat{\rho}} &=2\Im\Big\{ \langle\hat{S}^{(1)}_+(l)\hat{S}^{(2)}_+(l)\rangle_{\hat{\rho}}\nonumber\\
    &\quad- \langle\hat{S}^{(1)}_+(l)\hat{S}^{(2)}_-(l)\rangle_{\hat{\rho}}\Big\} \,.
\end{align}
We know both the terms on r.h.s from Eqs.~\eqref{eq:correlationS+S+2} and \eqref{eq:correlationS+S-2}. Turns out that correlation functions in \eqref{eq:correlationSzSx}, \eqref{eq:correlationSzSy} and \eqref{eq:correlationSxSy} are all 0 for a TMS vacuum state (proof is in Appx. A of Ref.~\cite{Main_paper}), i.e., measurements along orthogonal directions are uncorrelated for TMS vacuum state. 
Let us calculate the two-point correlation function for the spin-$y$ operator. Expanding the $\hat{S}_y(l)$ operator, the correlation function is found to be
\begin{align}
    \langle\hat{S}^{(1)}_y(l)\hat{S}^{(2)}_y(l)\rangle_{\hat{\rho}} &= 2\Re\Big\{ \langle\hat{S}^{(1)}_+(l)\hat{S}^{(2)}_-(l)\rangle_{\hat{\rho}} \nonumber\\
    &\quad-  \langle\hat{S}^{(1)}_+(l)\hat{S}^{(2)}_+(l)\rangle_{\hat{\rho}}\Big\}\,.
\end{align}
Again, we have calculated both the terms on r.h.s.~in Eqs.~\eqref{eq:correlationS+S+} and \eqref{eq:correlationS+S-}.  

In order to calculate the full expressions of the correlation functions for the TMS thermal state and perform the integration over the quadratures, we shall refer to the next appendix. 

\section{Covariance matrix and Wigner function for a TMS thermal state}\label{appx:cov_and_Wig}
In this appendix, we will compute the covariance matrix for a TMS thermal state and use it to evaluate the corresponding Wigner function and logarithmic negativity. 

We can quantify the first and second statistical moments of a quantum state in a compact way by building the operator-valued vector, $\boldsymbol{\hat{\xi}}\coloneqq(\hat{q}_1,\dots,\hat{q}_n,\hat{p}_1,\dots,\hat{p}_n)^T$. 
    The first statistical moment of the quantum state $\hat{\rho}$ is captured in the displacement vector $\boldsymbol{\bar{\xi}}$ which is given by\footnote{Note that since $\boldsymbol{\hat{\xi}}$ is an operator-valued \textit{vector}, we will denote its components with an upper index. For example, $\hat{\xi}^{1} = \hat{q}_1$.} 
    \begin{equation}
        \bar{\xi}^{\mu} = \langle\hat{\xi}^{\mu}\rangle_{\hat{\rho}}\,.
    \end{equation}
    The second moments are collected in the covariance matrix which is 
    \begin{equation}\label{eq:cov_mat}
        \sigma^{\mu\nu} = \langle\hat{\xi}^{\mu}\hat{\xi}^{\nu}+ \hat{\xi}^{\nu}\hat{\xi}^{\mu}\rangle_{\hat{\rho}}-2\langle\hat{\xi}^{\mu}\rangle_{\hat{\rho}}\langle\hat{\xi}^{\nu}\rangle_{\hat{\rho}}.
    \end{equation}

A Gaussian state is a quantum state fully characterized by its first two moments \cite{holevo}. In fact, the Wigner function for a Gaussian state can be expressed in terms of the first two moments as follows \cite{Eduardo_lecture_notes}:
    \begin{equation}\label{eq:Wigner_cov}
     W_{\hat{\rho}}(\boldsymbol{{\xi}}) = \frac{1}{\pi^n\sqrt{\text{det}(\boldsymbol{\sigma})}}\exp{-(\boldsymbol{\xi}-\boldsymbol{\bar\xi})^\mu(\boldsymbol{\xi}-\boldsymbol{\bar\xi})^\nu(\boldsymbol{\sigma}^{-1})_{\mu\nu}}\,.
    \end{equation}

The action of a Gaussian unitary $\hat{U}$ -- a unitary operation generated by a quadratic Hamiltonian ($\hat{U} = e^{-\ii \hat{H}}$) -- on a Gaussian state preserves its Gaussianity \cite{Gaussianity_preserve}. So, if we have a Gaussian state $\hat{\rho}$, then the state after the action of the Gaussian unitary given by 
\begin{equation}\label{eq:state_trans_gauss}
    \hat{\rho}' = \hat{U}\hat{\rho}\hat{U}^{\dagger},
\end{equation}
is also a Gaussian state. The corresponding transformation of the Wigner function in phase space is called a symplectic transformation, brought about by the symplectic transformation matrix $\mathbf{S}$. 

Let us now briefly outline the derivation of how the generating Hamiltonian $\hat{H}$ in Hilbert space is related to the symplectic transformation matrix $\mathbf{S}$ in phase space. For more details, the reader is encouraged to refer to (among many other references, \cite{Eduardo_lecture_notes}). 

In general, a time-independent Hamiltonian with only quadratic terms can be written as follows:
\begin{equation}\label{eq:H_in_F}
    \hat{H} = \frac{1}{2}\hat{\boldsymbol{\xi}}^{\text{T}}\mathbf{F}\hat{\boldsymbol{\xi}}\,,
\end{equation}
where $\mathbf{F}$ is a Hermitian matrix that captures coefficients of the terms appearing in the Hamiltonian. Let us express the full Hamiltonian $\hat{H}$ in terms of the creation and annihilation operators. We introduce $n$-dimensional vectors containing the creation and annihilation operators: 
$$\mathbf{\hat{a}} \coloneqq (\hat{a}_1,\hat{a}_2\dots,\hat{a}_n)^{\text{T}},\hspace{1cm}\mathbf{\hat{a}^{\dagger}} \coloneqq (\hat{a}^{\dagger}_1,\hat{a}^{\dagger}_2\dots,\hat{a}^{\dagger}_n)^{\text{T}}\,.$$
The Hamiltonian from Eq.~\eqref{eq:H_in_F} can then be written as
\begin{equation}\label{eq:H_in_crea_ann}
    \hat{H} = (\mathbf{\hat{a}^{\dagger}})^{\text{T}}\mathbf{W}\mathbf{\hat{a}} + (\mathbf{\hat{a}^{\dagger}})^{\text{T}}\mathbf{G}\mathbf{\hat{a}^{\dagger}}+ (\mathbf{\hat{a}})^{\text{T}}\mathbf{G^{\text{H}}}\mathbf{\hat{a}}\,,
\end{equation}
where $\mathbf{G}^{\text{H}}$ denotes the conjugate transpose of the matrix $\mathbf{G}$. Expanding the quadrature operators in Eq.~\eqref{eq:H_in_F} in terms of the creation and annihilation operators and comparing with Eq.~\eqref{eq:H_in_crea_ann}, one obtains $\mathbf{F}$ in the block form as 
\begin{equation}\label{eq:Fmat}
    \mathbf{F} = \begin{pmatrix}
        \mathbf{A}&\mathbf{X}\\
        \mathbf{X}^{\text{H}}&\mathbf{B}
    \end{pmatrix}\,,
\end{equation}
where 
\begin{align}
\mathbf{A}=\mathbf{W}+\mathbf{G}+\mathbf{G}^{\text{H}},&\phantom{=} \hspace{0.7cm} \mathbf{B}=\mathbf{W}-\mathbf{G}-\mathbf{G}^{\text{H}},\nonumber \\ \label{eq:A,B,X}\hspace{-7em}\mathbf{X}&=\ii(\mathbf{W}-\mathbf{G}+\mathbf{G}^{\text{H}})\,.
\end{align}
Then the symplectic transformation matrix $\mathbf{S}$ comes out to be \cite{Eduardo_lecture_notes}
\begin{equation}\label{eq:S_in_F}
    \mathbf{S} = \exp(\boldsymbol{\Omega}^{-1}\mathbf{\bar{F}})\,,
\end{equation}
where $\mathbf{\bar{F}} = \dfrac{\mathbf{F}+\mathbf{F}^{\text{T}}}{2}$ and $\boldsymbol{\Omega}$ is the symplectic matrix given by
\begin{equation}\label{eq:symplectic_matrix}
    \boldsymbol{\Omega} = \begin{pmatrix}
    0_n&-\openone_n\\
    \openone_n&0_n 
\end{pmatrix} \Rightarrow \boldsymbol{\Omega}^{-1} =\begin{pmatrix}
    0_n&\openone_n\\
    -\openone_n&0_n 
\end{pmatrix}\,.
\end{equation}
The action of Gaussian unitary in Hilbert space translates as a transformation of the covariance matrix in phase space by the symplectic transformation matrix, given by \cite{Eduardo_lecture_notes}
\begin{equation}\label{eq:cov_trans_symp}
    \boldsymbol{\sigma}' = \mathbf{S}\boldsymbol{\sigma}\mathbf{S}^{\text{T}}\,.
\end{equation}
To find the covariance matrix of a TMS thermal state, we first need to determine the symplectic matrix $\mathbf{S}_{\text{sq}}$ that implements squeezing in the phase space, by starting from the Hamiltonian generator of squeezing $\hat{H}_{\text{sq}}$ in Hilbert space. We will then apply the transformation \eqref{eq:cov_trans_symp} to the covariance matrix of a thermal state of two harmonic oscillators (which is also a Gaussian state).

The squeezing operator is a Gaussian unitary given by
\begin{equation}\label{eq:squeeze_op}
    \hat{S}(r,\phi) = \exp\left[r( e^{\ii\phi}\hat{a}^{\dagger}\hat{b}^{\dagger}-e^{-\ii\phi}\hat{a}\hat{b})\right] \,,
\end{equation}
and generated by the Hamiltonian
\begin{equation}\label{eq:ham_sq}
    \hat{H}_{\text{sq}} = \ii r \left(e^{i\phi}\hat{a}^{\dagger}\hat{b}^{\dagger}- e^{-\ii\phi}\hat{a}\hat{b}  \right). 
\end{equation}
Comparing with Eq.~\eqref{eq:H_in_crea_ann}, $\mathbf{W}$ matrix is 0 and the $\mathbf{G}$ matrix is found to be
\begin{equation}
    \mathbf{G} = \frac{\ii re^{i\phi}}{2} \begin{pmatrix} 0 & 1 \\ 1 & 0 \end{pmatrix} = \frac{\ii re^{i\phi}}{2} \sigma_x.
\end{equation}
Using Eq.~\eqref{eq:A,B,X}, $\mathbf{A}$, $\mathbf{B}$ and $\mathbf{X}$ matrices take the form:  
\begin{align}
\mathbf{A} &= -r\sin(\phi) \sigma_x = \mathbf{B}, \quad \mathbf{X} = r\cos(\phi) \sigma_x.
\end{align}
From Eq.~\eqref{eq:Fmat}, $\textbf{F}$ is found to be:
\begin{equation}
\mathbf{F} = r \begin{pmatrix}
-\sin(\phi)\sigma_x & \cos(\phi)\sigma_x \\
\cos(\phi)\sigma_x & \sin(\phi)\sigma_x \\
\end{pmatrix} \Rightarrow \mathbf{\bar{F}} = \frac{\mathbf{F}+\mathbf{F}^\text{T}}{2} = \mathbf{F},
\end{equation}
From Eq.~\eqref{eq:S_in_F}, the symplectic transformation matrix that implements squeezing in phase space is given by
\begin{widetext}
\begin{align}
    \mathbf{S_{\text{sq}}} &= \exp(\mathbf{\Omega}^{-1}\mathbf{\bar{F}}) \nonumber\\
    \mathbf{S_{\text{sq}}}&= \begin{pmatrix}
\cosh r &\cos(\phi) \sinh r & 0 &\sin(\phi)\sinh(r)\\
\cos(\phi) \sinh r& \cosh r & \sin(\phi)\sinh(r) & 0 \\
0& \sin(\phi)\sinh(r)&\cosh r  &-\cos(\phi) \sinh r\\
\sin(\phi)\sinh(r)&0&-\cos(\phi) \sinh r&\cosh r\\
\end{pmatrix}
\end{align}
\end{widetext}
The covariance matrix of a thermal state of a harmonic oscillator with frequency $\omega_A$ at temperature $T$ is given by \cite{Eduardo_lecture_notes}
\begin{equation}\label{eq:cov_thermal1}
    \boldsymbol\sigma_A = \nu_A \openone,
\end{equation}
where $\nu_A = \coth{\left(\omega_A/2 T\right)}$. For two Harmonic oscillators, with the other one having frequency $\omega_B$, we can write the covariance matrix $\boldsymbol\sigma_{\text{T}}$ as \cite{Eduardo_lecture_notes}
\begin{align}\label{eq:cov_thermal2}
    \boldsymbol\sigma_{\text{T}}&= \boldsymbol\sigma_A \oplus \boldsymbol\sigma_B \nonumber\\
    \boldsymbol\sigma_{\text{T}} &= \begin{pmatrix}
        \nu_A&0&0&0\\
        0&\nu_B&0&0\\
        0&0&\nu_A&0\\
        0&0&0&\nu_B\\
    \end{pmatrix},
\end{align}
where $\nu_B = \coth{\left(\omega_B/2 T\right)}$. 
The covariance matrix for the TMS thermal state ($\boldsymbol\sigma_{\text{ST}}$) is found by using Eq.~\eqref{eq:cov_trans_symp}. 
\begin{widetext}
    \begin{align}
    \boldsymbol\sigma_{\text{ST}} &= \mathbf{S_{\text{sq}}} \boldsymbol\sigma_{\text{T}}\mathbf{S^{\text{T}}_{\text{sq}}}\nonumber\\
    \boldsymbol\sigma_{\text{ST}}&=\frac{1}{2} \begin{pmatrix}
\nu_- + \nu_+ \cosh(2r) & \nu_+ \cos(\phi)\sinh(2r) & 0 & \nu_+ \sin(\phi) \sinh(2r) \\
\nu_+ \cos(\phi) \sinh(2r) & -\nu_- + \nu_+ \cosh(2r) & \nu_+\sin(\phi) \sinh(2r) & 0 \\
0 & \nu_+\sin(\phi) \sinh(2r) & \nu_- + \nu_+\cosh(2r) & -\nu_+ \cos(\phi) \sinh(2r) \\
\nu_+\sin(\phi) \sinh(2r) & 0 & -\nu_+ \cos(\phi) \sinh(2r) & -\nu_- + \nu_+ \cosh(2r)
\end{pmatrix},
\end{align}
where $\nu_+ = \nu_A+\nu_B$ while $\nu_- = \nu_A-\nu_B$. If $\nu_a = \nu_B = \nu$, i.e., oscillators have the same frequency and represent $\boldsymbol\sigma_{\text{ST}}$ in $\boldsymbol{\hat{\xi}} =(\hat{q_1},\hat{p_1},\hat{q_2}, \hat{p_2})^{\text{T}}$ quadrature ordering, we get
\begin{equation}\label{eq:cov_TMST}
    \boldsymbol\sigma_{\text{ST}} = \nu\begin{pmatrix}
\cosh(2r) & 0 &\cos(\phi)\sinh(2r) & \sin(\phi) \sinh(2r) \\
0& \cosh(2r) & \sin(\phi) \sinh(2r)& -\cos(\phi) \sinh(2r)\\
\cos(\phi) \sinh(2r)& \sin(\phi) \sinh(2r) & \cosh(2r) & 0 \\
\sin(\phi) \sinh(2r)  & -\cos(\phi) \sinh(2r)&0 & \cosh(2r)
\end{pmatrix} .
\end{equation}
\end{widetext}

Substituting the above expression in Eq.~\eqref{eq:Wigner_cov}, we obtain
\begin{multline}\label{eq:Wig_TMST_appx}
   W_{\hat{\rho}_{\text{ST}}}(q_1,q_2,p_1,p_2) = \frac{1}{(\nu(T)\pi)^2}\\
    \times e^{\frac{1}{\nu(T)}[2\sinh(2r)(q_1q_2-p_1p_2) - \cosh{(2r)(q_1^2+q_2^2+p_1^2+p_2^2)} ]}, \raisetag{14mm}
\end{multline}
where $\nu(T)\equiv \nu=\coth{\left(\omega/2 T\right)}$. Note that the displacement vector for the TMS thermal state is 0, since the Hamiltonian of a harmonic oscillator and the generator of squeezing are both quadratic with no linear terms, which means if the displacement vector initially has a zero mean, it will remain zero even after the transformation \cite{Eduardo_paper_Gaussian_QM}.

Let us now proceed to compute the entanglement in a TMS thermal state using its covariance matrix. The most common measure of entanglement used for Gaussian states is logarithmic negativity $E_N$. It is derived from the Peres-Horodecki criterion \cite{Peres,Horodecki} for separability which says that if the partial transpose of the density operator $\hat{\rho}$ of a bipartite state has a negative eigenvalue, then the state is entangled. The negativity $\mathcal{N}$, then, is simply the absolute sum of the negative eigenvalues of the partial transpose, and the logarithmic negativity is given by
$$E_N = \log_2(2\mathcal{N}+1)\,.$$

For bipartite Gaussian states, there is an alternative way to compute $E_N$ from the covariance matrix \cite{Alessio_ent_cov}. In the $\boldsymbol{\hat{\xi}} = (\hat{q}_1,\hat{q}_2,\hat{p}_1,\hat{p}_2)$ quadrature ordering, the covariance matrix can be arranged in the block form as 
\begin{equation}\label{eq:cov_block}
    \boldsymbol{\sigma} = \begin{pmatrix}
        \boldsymbol{\sigma}_1&\boldsymbol{\sigma}_{12}\\
        \boldsymbol{\sigma}^{\text{T}}_{12}&\boldsymbol{\sigma}_2
    \end{pmatrix}\,,
\end{equation}
where $\boldsymbol{\sigma}_1$ and $\boldsymbol{\sigma}_2$ are the $2\cross2$ covariance matrices for subsystems 1 and 2, while $\boldsymbol{\sigma}_{12}$ is a matrix that provides correlations between the two subsystems. 

Let $\Delta$ be a number given by 
\begin{equation}
    \Delta = \text{det}(\boldsymbol{\sigma}_1) + \text{det}(\boldsymbol{\sigma}_2) - 2\text{det}(\boldsymbol{\sigma}_{12})\,.
\end{equation}
The partially transposed covariance matrix $\boldsymbol{\sigma}^{\text{PT}}$ can be obtained by flipping the sign of the momentum operator $\hat{p}_2$ in $\boldsymbol{\sigma}$. Then, the logarithmic negativity $E_N$ is given by
\begin{equation}\label{E_N}
    E_N = \text{max}(0,-\log{n_-}),
\end{equation} 
where $n_{\pm}$ are the eigenvalues of the matrix $\ii\boldsymbol{\Omega}^{-1}\boldsymbol{\sigma}^{\text{PT}}$ called the symplectic eigenvalues of the partially-transposed covariance matrix and are given by 
\begin{equation}\label{eq:nupm}
    n_{\pm} = \sqrt{\frac{\Delta\pm\sqrt{\Delta^2-4\text{det}(\boldsymbol{\sigma})}}{2}}\,.
\end{equation}

For the TMS thermal state, we can find the logarithmic negativity using Eqs.~\eqref{eq:cov_block}--\eqref{eq:nupm} by changing the quadrature ordering in Eq.~\eqref{eq:cov_TMST}. In that case, we have $\boldsymbol{\sigma}_1, \boldsymbol{\sigma}_2$ and $\boldsymbol{\sigma}_{12}$ matrices as
\begin{align*}
    \boldsymbol{\sigma_1} = \boldsymbol{\sigma_2} &= \nu\begin{pmatrix}
    \cosh(2r) & 0 \\
    0 & \cosh(2r)
\end{pmatrix} \\ \boldsymbol{\sigma_{12}} &= \nu\begin{pmatrix}
    \cos(\phi)\sinh(2r) & \sin(\phi) \sinh(2r)\\
    \sin(\phi) \sinh(2r)& -\cos(\phi) \sinh(2r)
\end{pmatrix}\,.
\end{align*}
Thus, using the above expressions, we have numerically found the logarithmic negativity for the TMS thermal state and plotted the same as a function of squeezing and temperature in Fig.~\ref{fig:logN_r_T}.
\section{Calculating the correlation functions for the TMS thermal state}\label{Appendix:Correlation_funcs_TMST}
In this appendix, we will evaluate the two-point correlation functions of the spin operators, particularly the spin $z-z$ and spin $x-x$ correlators, by substituting the Wigner representation of the TMS thermal state given in Eq.~\eqref{eq:Wigner_TMS_thermal}. The calculations that follow proceed along the same lines as those presented in Appx. A of Ref.~\cite{Main_paper}.

\subsection{$\langle\hat{S}^{(1)}_z(l)\hat{S}^{(2)}_z(l)\rangle_{\hat{\rho}_{\text{ST}}}$ correlation function}
Let us start with $\langle\hat{S}^{(1)}_z(l)\hat{S}^{(2)}_z(l)\rangle_{\hat{\rho}_{\text{ST}}}$. From Eqs.~\eqref{eq:correlationSzSz2} and \eqref{eq:Wig_TMST_appx}, we have 
\begin{align}
\left\langle\hat{S}^{(1)}_z(l)\hat{S}^{(2)}_z(l)\right\rangle_{\hat{\rho}_{\text{ST}}} &\!\!\!=\frac{1}{\pi}\!\!\sum_{m,n=-\infty}^{\infty}\!\!\!(-1)^{m+n}\!\!\!\int_{nl}\limits^{(n+1)l} \!\!\!\!\dd{q_1} \int_{ml}\limits^{(m+1)l} \!\!\!\!\dd{q_2}\nonumber\\ &\hspace{-6em}\phantom{=}\times 
\frac{1}{(\nu(T))^{3/2}}e^{\frac{1}{\nu(T)}(-(q_1^2+q_2^2)\cosh{(2r)} + 2q_1q_2\sinh(2r))}\nonumber \\
&=\frac{1}{\pi}\sum_{m,n=-\infty}^{\infty}(-1)^{m+n} Z_{n,m}\,.
\end{align}
We solve for $Z_{n,m}$ from here. We notice that the integral resembles a Gaussian integral.~To make it explicit, we substitute $q_1 = u+v$ and $q_2=u-v$. Then, the exponential becomes 
\begin{align}
    &\quad \quad e^{\frac{1}{\nu(T)}(-2(u^2+v^2)\cosh{(2r)} + 2(u^2-v^2)\sinh(2r))}, \nonumber\\
    &\Rightarrow e^{\frac{1}{\nu(T)}(-2u^2(\cosh{(2r)} - \sinh{(2r)}) - 2v^2\cosh{(2r)}+\sinh(2r)})\,. \nonumber
\end{align}
Let us define 
\begin{equation}
\label{eq:gamma1}
    \gamma_1(r,T)\equiv\gamma_1 = \tfrac{2}{\nu(T)}(\cosh{(2r)} - \sinh{(2r)}) = \tfrac{2}{\nu(T)}e^{-2r}
\end{equation}
\begin{equation}
\label{eq:gamma2}
    \gamma_2(r,T)\equiv\gamma_2 = \tfrac{2}{\nu(T)}(\cosh{(2r)} + \sinh{(2r)}) = \tfrac{2}{\nu(T)}e^{2r}\,.
\end{equation}
With this, the exponential turns into two Gaussian functions in $u$ and $v$. And so the double integration can be expressed as a product of two one-dimensional integrals. The integration in $Z_{n,m}$ with the appropriate limits looks like 
\begin{align}
Z_{n,m} &= \frac{2}{(\nu(T))^{3/2}}\Bigg[\int_{(n+m)\tfrac{l}{2}}\limits^{(n+m+1)\tfrac{l}{2}}\!\!\!\!\dd{u}e^{-\gamma_1 u^2}\int_{nl-u}\limits^{u-ml}\dd{v}e^{-\gamma_2v^2} 
\nonumber\\ &\phantom{=}+\int_{(n+m+1)\tfrac{l}{2}}\limits^{(n+m+2)\tfrac{l}{2}}\!\!\!\!\dd{u}e^{-\gamma_1 u^2}\int_{u-(m+1)l}\limits^{(n+1)l-u}\dd{v}e^{-\gamma_2v^2}\Bigg]\nonumber\\
&= Z^{(1)}_{n,m} + Z^{(2)}_{n,m}\,,
\end{align}
where the factor of 2 in front comes from the Jacobian. Performing the integral over $v$ in $Z^{(1)}_{n,m}$, we get
\begin{align}
    Z^{(1)}_{n,m} &= \frac{2}{(\nu(T))^{3/2}}\int_{(n+m)\tfrac{l}{2}}\limits^{(n+m+1)\tfrac{l}{2}}\dd{u}e^{-\gamma_1 u^2}\left(\tfrac{1}{2}\sqrt{\tfrac{\pi}{\gamma_2}}\right)\nonumber\\ &\phantom{=}\times\left[\erf\left((u-ml)\sqrt{\gamma_2}\right) + \erf\left((u-nl)\sqrt{\gamma_2}\right)\right]\,.\nonumber
\end{align}
Substituting $u = (z+n+m)\tfrac{l}{2}$
\begin{align}
    Z^{(1)}_{n,m} &= \frac{l}{2(\nu(T))^{3/2}}\sqrt{\frac{\pi}{\gamma_2}} \int_0^1 \dd{z} e^{-\gamma_1(z+n+m)^2\tfrac{l^2}{4}}\times\nonumber\\ &\hspace{-4em}\phantom{=}\left[\erf\left(\tfrac{l}{2}\sqrt{\gamma_2}(z+n-m)\right)+ \erf\left(\tfrac{l}{2}\sqrt{\gamma_2}(z-n+m)\right)\right].
\end{align}
Similar result is obtained for $Z^{(2)}_{n,m}$ by substituting $u = (z-n-m-2)\tfrac{l}{2}$ after integrating over $v$:
\begin{align}
    Z^{(2)}_{n,m} &= \frac{l}{2(\nu(T))^{3/2}}\sqrt{\frac{\pi}{\gamma_2}} \int_0^1 \dd{z} e^{-\gamma_1(z-n-m-2)^2\tfrac{l^2}{4}} \times\nonumber\\ \label{eq:Z_nm_appx}&\hspace{-4em}\phantom{=}\left[\erf\left(\tfrac{l}{2}\sqrt{\gamma_2}(z+n-m)\right)+ \erf\left(\tfrac{l}{2}\sqrt{\gamma_2}(z-n+m)\right)\right]\\
        &= Z^{(1)}_{-n-1,-m-1}\,.
\end{align}
Since $Z^{(2)}_{n,m}$ can be expressed in terms of $Z^{(1)}_{n,m}$ and due to the symmetry of the summands $n$ and $m$ which run from $-\infty$ to $\infty$, $\sum_{n,m}Z^{(2)}_{n,m} = \sum_{n,m}Z^{(1)}_{n,m} \equiv \frac{1}{2}\sum_{n,m}Z_{n,m}$. Thus, the correlation function $\langle\hat{S}^{(1)}_z(l)\hat{S}^{(2)}_z(l)\rangle_{\hat{\rho}_{\text{ST}}}$ is found to be:
\begin{equation}
\left\langle\hat{S}^{(1)}_z(l)\hat{S}^{(2)}_z(l)\right\rangle_{\hat{\rho}_{\text{ST}}} = \frac{2}{\pi}\sum_{m,n=-\infty}^{\infty}(-1)^{m+n} Z^{(1)}_{n,m}\,.
\end{equation}

\subsection{$\langle\hat{S}^{(1)}_x(l)\hat{S}^{(2)}_x(l)\rangle_{\hat{\rho}_{\text{ST}}}$ correlation function}
The evaluation of this correlation function includes calculating both $\langle\hat{S}^{(1)}_+(l)\hat{S}^{(2)}_+(l)\rangle_{\hat{\rho}_{\text{ST}}}$ and $\langle\hat{S}^{(1)}_+(l)\hat{S}^{(2)}_-(l)\rangle_{\hat{\rho}_{\text{ST}}}$. Let us start with the former. From Eq.~\eqref{eq:correlationS+S+2}, we have 
\begin{align}
    \left\langle\hat{S}^{(1)}_+(l)\hat{S}^{(2)}_+(l)\right\rangle_{\hat{\rho}_{\text{ST}}} &\!\!\!= \!\!\!\!\sum_{m,n=-\infty}^{\infty} \int_{-\infty}^{\infty}\!\!\!\!\dd{p_1}\dd{p_2}\!\!\!\! \int_{2nl}\limits^{(2n+1)l} \!\!\!\!\dd{q_1} \int_{2ml}\limits^{(2m+1)l} \!\!\!\!\!\!\!\!\dd{q_2}\nonumber\\ &\hspace{-5em}\phantom{=}\times  e^{\ii(p_1+p_2)l}\, W_{\hat{\rho}_{\text{ST}}}(q_1+\tfrac{l}{2},q_2+\tfrac{l}{2},p_1,p_2) \\
    &\!\!\!= \frac{1}{\pi}\sum_{m,n=-\infty}^{\infty} J_{n,m}\,, \nonumber
\end{align}
Integrating the exponential factor and the Wigner function given in Eq.~\eqref{eq:Wig_TMST_appx} over $p_1$ and $p_2$, $J_{n,m}$ comes out to be:
\begin{align}
  J_{n,m} &=\!\!\!\!\! \int_{2nl}\limits^{(2n+1)l} \!\!\!\!\dd{q_1} \!\!\!\int_{2ml}\limits^{(2m+1)l}\!\!\!\! \dd{q_2} \frac{1}{\nu(T)}e^{-l^2\nu\left(\frac{T}{2}\right)(\cosh(2r) - \sinh(2r))}\times\nonumber \\
   &\hspace{-3em}\phantom{={}}  \!e^{\frac{-1}{\nu(T)}\!\{(q_1^2+q_2^2)\cosh(2r) - 2q_1q_2\sinh(2r)+(\cosh(2r)-\sinh(2r))l(q_1+q_2)\}}\!. 
\end{align}
We note that as $T\rightarrow 0$, $\nu(T)\rightarrow 1$ and so does $\nu\left(\tfrac{T}{2}\right)$. In that case, the above equation looks like
\begin{align}\label{eq:Jnm}
    J_{n,m} &= \!\!\!\int_{2nl}\limits^{(2n+1)l} \!\!\!\!\dd{q_1} \int_{2ml}\limits^{(2m+1)l}\!\!\!\! \dd{q_2} e^{-l^2(\cosh(2r)-\sinh(2r))}\times\nonumber\\ &\hspace{-3em}\phantom{=} e^{-\left\{(q_1^2+q_2^2)\cosh(2r) - 2q_1q_2\sinh(2r)+(\cosh(2r)-\sinh(2r))l(q_1+q_2)\right\}}.
\end{align}
This matches exactly the expression for $J_{n,m}$ for a TMS vacuum state given in Appendix A of Ref.~\cite{Main_paper}. We repeat the same steps for $\langle\hat{S}^{(1)}_+(l)\hat{S}^{(2)}_-(l)\rangle_{\hat{\rho}_{\text{ST}}}$. From Eq.~\eqref{eq:correlationS+S-}, 
\begin{align}
     \left\langle\hat{S}^{(1)}_+(l)\hat{S}^{(2)}_-(l)\right\rangle_{\hat{\rho}_{\text{ST}}} &\!\!=\!\!\! \sum_{m,n=-\infty}^{\infty} \int_{-\infty}^{\infty}\!\!\!\!\dd{p_1}\dd{p_2} \!\!\!\!\!\int_{2nl}\limits^{(2n+1)l}\!\!\!\!\! \dd{q_1} \!\int_{(2m+1)l}\limits^{(2m+2)l}\!\!\!\!\!\! \dd{q_2}\nonumber\\
     &\hspace{-4.5em}\phantom{=}\times e^{\ii(p_1-p_2)l}\, W(q_1+\tfrac{l}{2},q_2-\tfrac{l}{2},p_1,p_2)\\
    &\!\!= \frac{1}{\pi}\sum_{m,n=-\infty}^{\infty} K_{n,m}\,, \nonumber
\end{align}
Integrating over $p_1$ and $p_2$, we get 
\begin{align}
    K_{n,m} &\!= \!\!\!\!\!\!\int_{2nl}\limits^{(2n+1)l}\!\!\!\!\!\! \dd{q_1} \!\!\!\int_{(2m+1)l}\limits^{(2m+2)l} \!\!\!\!\!\frac{\dd{q_2}}{\nu(T)}e^{-l^2\nu\left(\frac{T}{2}\right)(\cosh(2r)+\sinh(2r)}\nonumber \\
    &\hspace{-3.5em}\phantom{={}}\times  e^{\frac{-1}{\nu(T)}\{(q_1^2+q_2^2+l(q_1-q_2))\cosh(2r) - (2q_1q_2-l(q_1-q_2))\sinh(2r)\}}\,.
\end{align}
To evaluate the spin-$x$ two-point correlation function, we need to add both $J_{n,m}$ and $K_{n,m}$. We note that the integration limits for the $q_2$ variable are different in both. To align these limits, we do the following substitution: $q_1 = y_1$ and $q_2 = y_2+l$ in $K_{n,m}$. This yields
\begin{align}
    K_{n,m} &\!\!=\!\!\!\!\!\!\!\!\int_{2nl}\limits^{(2n+1)l}\!\!\!\!\!\!\!\! \dd{y_1} \!\!\int_{2ml}\limits^{(2m+1)l}\!\!\!\!\!\!\! \frac{\dd{y_2}}{\nu(T)} e^{\!\frac{l^2}{\nu(T)}\sinh(2r)-l^2\nu\left(\frac{T}{2}\right)(\cosh(2r)+\sinh(2r))} \nonumber \\ \label{eq:Knm}
    &\hspace{-3.8em}\phantom{={}}\times e^{\frac{-1} {\nu(T)}\{(y_1^2+y_2^2+l(y_1+y_2))\cosh(2r) - (2y_1y_2 +l(y_1+y_2))\sinh(2r)\}}.
\end{align}
Thus, we can write the spin-$x$ correlation function by adding Eqs.~\eqref{eq:Jnm} and \eqref{eq:Knm}. From Eq.~\eqref{eq:correlationSxSx}, we have 
\begin{align}
    \left\langle\hat{S}^{(1)}_x(l)\hat{S}^{(2)}_x(l)\right\rangle_{\hat{\rho}_{\text{ST}}} &\!\!\!= 2\Re\Big\{\left\langle\hat{S}^{(1)}_+(l)\hat{S}^{(2)}_+(l)\right\rangle_{\hat{\rho}_{\text{ST}}}\nonumber\\&\phantom{=}+ \left\langle\hat{S}^{(1)}_+(l)\hat{S}^{(2)}_-(l)\right\rangle_{\hat{\rho}_{\text{ST}}}\Big\}\nonumber\\
    \Rightarrow\left\langle\hat{S}^{(1)}_x(l)\hat{S}^{(2)}_x(l)\right\rangle_{\hat{\rho}_{\text{ST}}} &\!\!\!= \frac{2}{\pi}\Re\Big\{\sum_{m,n=-\infty}^{\infty}(J_{n,m}+K_{n,m})\Big\} \nonumber \,.
\end{align}
For a vanishing squeezing angle, both $J_{n,m}$ and $K_{n,m}$ are real. Thus,
\begin{equation}
    \left\langle\hat{S}^{(1)}_x(l)\hat{S}^{(2)}_x(l)\right\rangle_{\hat{\rho}_{\text{ST}}} = \frac{2}{\pi}\sum_{m,n=-\infty}^{\infty}(J_{n,m}+K_{n,m})\,.
\end{equation}
Before adding the Eqs.~\eqref{eq:Jnm} and \eqref{eq:Knm}, we rename the variables $q_1$ and $q_2$ in \eqref{eq:Jnm} to $y_1$ and $y_2$. Note that the exponential in the first line of both equations does not depend on the integration variables and therefore can be factored out of the integration. We will collect those exponential terms and incorporate them into a function named $\Gamma(l,r,T)$.
\begin{align}\label{eq:D.15}
    J_{n,m}+K_{n,m} &\!\!=\!\! \frac{\Gamma(l,r,T)}{\nu(T)}\!\!\!\!\!\int_{2nl}\limits^{(2n+1)l} \!\!\!\!\!\!\dd{y_1} \!\!\!\int_{2ml}\limits^{(2m+1)l} \!\!\!\!\!\!\dd{y_2}e^{\frac{2}{\nu(T)}y_1y_2\sinh(2r)} \nonumber\\
    &\hspace{-6.8em}\phantom{={}} \times \! e^{\frac{-1} {\nu(T)}\{(y_1^2+y_2^2)\cosh(2r)+(\cosh(2r)-\sinh(2r))l(y_1+y_2)\}}\!,
\end{align}
where 
$$\Gamma(l,r,T) = e^{-2l^2\nu\left(\frac{T}{2}\right)e^{-2r}}+e^{-l^2\left\{2\nu\left(\frac{T}{2}\right)e^{2r}-\frac{1}{\nu(T)}\sinh(2r)\right\}}\,.$$
Let us substitute $\gamma_1$ and $\gamma_2$ in place of $\cosh(2r)$ and $\sinh(2r)$ in the exponential in Eq.~\eqref{eq:D.15}. The exponential becomes
\begin{align}
    \Rightarrow e^{-\left[\left(\frac{\gamma_1+\gamma_2}{4}\right)(y_1^2+y_2^2) + \left(\frac{\gamma_1- \gamma_2}{2}\right)y_1y_2
    +\frac{\gamma_1}{2}l(y_1+y_2)\right]}\,,
\end{align}
As before, in order to factorize it into two Gaussian integrals, we make the following substitution: $y_1 = u+v$ and $y_2 = u-v$. The exponential becomes
$$\Rightarrow \exp{-\gamma_1 u^2-\gamma_1 l u - \gamma_2 v^2}\,.$$
Reinstating the appropriate limits (much like before), we get
\begin{widetext}
   \begin{align}
    J_{n,m} + K_{n,m} &= \frac{2}{\nu(T)}\Gamma(l,r,T)\!\!\!\!\!\!\!\!\!\int_{(n+m)l}\limits^{(2n+2m+1)\tfrac{l}{2}}\!\!\!\!\!\!\!\!\!\dd{u}e^{-\gamma_1 u^2-\gamma_1 l u}\int_{2nl-u}\limits^{u-2ml}\!\!\!\!\dd{v}e^{-\gamma_2v^2} +\frac{2}{\nu(T)}\Gamma(l,r,T)\!\!\!\!\!\!\!\!\!\int_{(2n+2m+1)\tfrac{l}{2}}\limits^{(n+m+1)l}\!\!\!\!\!\!\!\!\!\dd{u}e^{-\gamma_1 u^2-\gamma_1 lu}\int_{u-(2m+1)l}\limits^{(2n+1)l-u}\!\!\!\!\!\!\!\dd{v}e^{-\gamma_2v^2} \nonumber\\
    &= X^{(1)}_{n,m}+ X^{(2)}_{n,m} \,,
\end{align} 
\end{widetext}
where the factor of 2 is from the Jacobian. The initial double integration has been expressed as a product of two one-dimensional Gaussian integrals as intended. We repeat the same steps as we did for $Z^{(1)}_{n,m}$ while solving for the spin-$z$ two-point correlation function. We integrate over $v$ and then substitute $u = (z+2n+2m)\tfrac{l}{2}$. This results in
\begin{align}\label{eq:Xnm1}
    X^{(1)}_{n,m} &\!\!=\!\!\frac{l\,\Gamma(l,r,T)}{2\nu(T)}\sqrt{\frac{\pi}{\gamma_2}} \int_0^1 \!\!\!\!\dd{z} e^{-\gamma_1(z+2n+2m)\left(z+2n+2m+2\right)\tfrac{l^2}{4}}\nonumber \\
        &\hspace{-3.5em}\phantom{={}} \times \!\!\left[\erf\left(\tfrac{l}{2}\sqrt{\gamma_2}(z+2n-2m)\right)\!+ \!\erf\left(\tfrac{l}{2}\sqrt{\gamma_2}(z-2n+2m)\right)\right].
\end{align}
We get a similar expression for $X^{(2)}_{n,m}$ by substituting $u = -(z-2n-2m-2)\tfrac{l}{2}$ after integrating over $v$:
\begin{align}
    X^{(2)}_{n,m} &\!\!=\!\!\frac{l\,\Gamma(l,r,T)}{2\nu(T)}\!\!\sqrt{\frac{\pi}{\gamma_2}}\!\! \int_0^1 \!\!\!\!\!\dd{z} e^{-\gamma_1(z-2n-2m-2)\left(z-2n-2m-4\right)\tfrac{l^2}{4}}\nonumber \\
    &\hspace{-3.5em}\phantom{={}} \times\!\! \left[\erf\left(\tfrac{l}{2}\sqrt{\gamma_2}(z+2n-2m)\right)\!+ \!\erf\left(\tfrac{l}{2}\sqrt{\gamma_2}(z-2n+2m)\right)\right]\\
    &\!\!\!=X^{(1)}_{-n-1,-m-1}\,.
\end{align}
Since, $X^{(2)}_{n,m}$ can be written in terms of $X^{(1)}_{n,m}$ and due to the symmetry of the summation, we can conclude that $\sum_{m,n=-\infty}^{\infty}X^{(2)}_{n,m} = \sum_{m,n=-\infty}^{\infty}X^{(1)}_{n,m}\equiv \frac{1}{2}\sum_{m,n=-\infty}^{\infty}X_{n,m}$. This means that 
$$\sum_{m,n=-\infty}^{\infty}(J_{n,m} + K_{n,m}) = 2\sum_{m,n=-\infty}^{\infty}X^{(1)}_{n,m}\,.$$
Thus finally, we get 
\begin{equation}
     \left\langle\hat{S}^{(1)}_x(l)\hat{S}^{(2)}_x(l)\right\rangle_{\hat{\rho}_{\text{ST}}} \!\!\!= \frac{4}{\pi}\sum_{m,n=-\infty}^{\infty}\!\!\!
     X^{(1)}_{n,m} = \frac{2}{\pi}\sum_{m,n=-\infty}^{\infty}\!\!\!X_{n,m} \,,
\end{equation}
with $X^{(1)}_{n,m}$ given by Eq.~\eqref{eq:Xnm1}.

Also, from Eqs.~\eqref{eq:Z_nm_appx} and \eqref{eq:Xnm1}, we can see that the temperature dependence factors out in terms of $\nu(T)$ from the expressions. Thus, following the same derivation for the correlation functions in orthogonal directions (such as $\langle\hat{S}^{(1)}_x(l)\hat{S}^{(2)}_z(l)\rangle$) as given in Appx. A of Ref.~\cite{Main_paper}, we find that these vanish for the TMS thermal state, just as they do for the TMS vacuum state.
\section{Large and small $l$ limit}
\label{appx:l_limit}
In Appendix B of Ref.~\cite{Main_paper}, the large and small $l$ limits for the correlation functions $\langle\hat{S}^{(1)}_z(l)\hat{S}^{(2)}_z(l)\rangle_{\hat{\rho}_{\text{SV}}}$ and $\langle\hat{S}^{(1)}_x(l)\hat{S}^{(2)}_x(l)\rangle_{\hat{\rho}_{\text{SV}}}$ for the TMS vacuum state have been discussed in much detail. The only difference in our case is the inclusion of temperature dependence in the expressions. However, since that does not affect the outcomes when taking the limits of large or small $l$, we will be presenting the results in this section without further elaboration. The entirety of the calculation follows exactly as in Ref.~\cite{Main_paper}.
\begin{enumerate}
    \item In the large $l$ limit, the spin-z correlation function becomes
    \begin{equation}
    \left\langle\hat{S}^{(1)}_z(l)\hat{S}^{(2)}_z(l)\right\rangle_{\hat{\rho}_{\text{ST}}} \!\!\!\!\!\!\simeq \frac{2}{\pi\sqrt{\nu(T)}} \arctan(\sinh(2r)) \,.
    \end{equation}
Also note that as $T\rightarrow0$, $\nu(T)\rightarrow1$ and in this limit, the above equation matches the expression given in Ref.~\cite{Main_paper}. This also coincides with Eq.~\eqref{eq:Bell_Gour} which is the expectation value of the spin-$x$ (unbinned) pseudospin operator \eqref{eq:pseudo_pos} for the TMS vacuum state.

$\langle\hat{S}^{(1)}_x(l)\hat{S}^{(2)}_x(l)\rangle_{\hat{\rho}_{\text{SV}}}$ (and also $\langle\hat{S}^{(1)}_y(l)\hat{S}^{(2)}_y(l)\rangle_{\hat{\rho}_{\text{SV}}}$) correlation function turns out to be $\simeq 0$ in the large $l$ limit \cite{Main_paper}. The proof presented in \cite{Main_paper} applies to the TMS vacuum state; however, a similar argument can be made for the TMS thermal state. This is because $\nu(T)$ containing the temperature dependence simply factors out in all relevant expressions. Thus, from Eq.~\eqref{eq:Bell_op}, the Bell operator expectation value turns out to be
\begin{equation}
    \langle\hat{B}\rangle_{\hat{\rho}_{\text{ST}}} = \frac{4}{\pi\sqrt{\nu(T)}} \arctan(\sinh(2r)) \,.
\end{equation}
Indeed, the asymptotes at large $l$ values in Figs.~\ref{fig:B_l}a and~\ref{fig:B_l}b match precisely with the results obtained by plugging in the corresponding values of $r$ and $T$ in the above equation.
\item In the small $l$ limit, we obtain
\begin{equation}
\left\langle\hat{S}^{(1)}_z(l)\hat{S}^{(2)}_z(l)\right\rangle_{\hat{\rho}_{\text{ST}}} \simeq \frac{2}{\sqrt{\nu(T)}} e^{-\tfrac{\pi^2\nu(T)e^{-2r}}{2l^2}}\,,
\end{equation}
which vanishes as $l\rightarrow 0$. For $\langle\hat{S}^{(1)}_x(l)\hat{S}^{(2)}_x(l)\rangle_{\hat{\rho}_{\text{ST}}} $, we have the factor of $\Gamma(l,r,T)$ from Eq.~\eqref{eq:Xnm1} which approaches 2 as $l\rightarrow 0$. The remainder of the calculation matches exactly and it is found that 
$$\left\langle\hat{S}^{(1)}_x(l)\hat{S}^{(2)}_x(l)\right\rangle_{\hat{\rho}_{\text{ST}}} \simeq \frac{\Gamma(l,r,T)}{2}\simeq 1 \,,$$
Thus, in the small $l$ limit, $\langle\hat{B}\rangle \simeq 2$. One can argue that this is not what is observed in Fig.~\ref{fig:B_l} where the asymptotes seem to be below 2 at small $l$ for larger values of $r$ and $T$. It turns out that for larger $r$ and $T$ values, the Bell curve converges to 2 at even smaller values of $l$. This is because some terms like $\exp{-2l^2\nu\left(\tfrac{T}{2}\right)e^{-2r}}$ (as in Eq.~\eqref{eq:Gamma}) has a factor of $\nu\left(\tfrac{T}{2}\right)$ in the exponential which grows rapidly as $T$ increases. Consequently, lower $l$ values are needed for the approximation to remain valid.
\end{enumerate}

\section{Correlation functions for (unbinned) pseudospin operators in Wigner representation}\label{appendix:unbinned_Bell_op}
In this appendix, we will follow a calculation similar to that outlined in Appx.~\ref{appx:correlation_funcs} and \ref{Appendix:Correlation_funcs_TMST} but for the (unbinned) pseudospin operators given in Eq.~\eqref{eq:pseudo_pos}. We will express the correlation functions $\langle \hat{\Pi}_z^{(1)}\hat{\Pi}_z^{(2)}\rangle_{\hat{\rho}}$ and $\langle \hat{\Pi}_x^{(1)}\hat{\Pi}_x^{(2)}\rangle_{\hat{\rho}}$ in the Wigner representation and then use those expressions to compute $\langle\hat{B}\rangle_{\hat{\rho}_{\text{ST}}}$ from Eq.~\eqref{eq:Bell_op}. 

Inserting $\hat{\Pi}_z^{(i)}$ from Eq.~\eqref{eq:pseudo_pos} in Eq.\eqref{eq:Wig_trans}, we get:
\begin{align}
    \Pi_z^{(i)}(q_i,p_i)\!&=\!-\!\!\int_{-\infty}^{\infty} \!\!\!\!\!\!\!\dd{x} \int_{-\infty}^{\infty}\!\!\!\!\!\!\! \dd{q} e^{-\ii p_i x}\langle q_i+\frac{1}{2}x|q\rangle\langle -q|q_i-\frac{1}{2}x\rangle\nonumber\\ \label{eq:Pizi}
    &=-\int_{-\infty}^{\infty} \dd{q} e^{-2\ii p_i(q-q_i)}\delta(q_i)\,.
\end{align}
One can obtain \eqref{eq:Pizi} simply following the steps given in Eq.~\eqref{eq:step1} through Eq.~\eqref{eq:Sz1}. The spin-$z$ correlation function for a state $\hat{\rho}$ is then given by 
\begin{align}
    \langle \hat{\Pi}_z^{(1)}\hat{\Pi}_z^{(2)}\rangle_{\hat{\rho}} &= \!\!\int_{-\infty}^{\infty}\!\!\!\!\dd{q_1}\dd{q_2}\dd{p_1}\dd{p_2}\Pi_z^{(1)}(q_1,p_1)\Pi_z^{(2)}(q_2,p_2)\nonumber\\
    &\quad\times W_{\hat{\rho}}(q_1,q_2,p_1,p_2)\\
    &\hspace{-5.5em}= \!\!\int_{-\infty}^{\infty}\!\!\!\!\dd{q_1}\dd{q_2}\dd{p_1}\dd{p_2} e^{-2\ii(p_1q_1+p_2q_2)}W_{\hat{\rho}}(0,0,p_1,p_2).
\end{align}
For the TMS vacuum state, this comes out to be 1. But for the TMS thermal state, we can substitute $W_{\hat{\rho}_{\text{ST}}}$ from Eq.~\eqref{eq:Wig_TMST_appx} and integrate over $q_i$'s and $p_i$'s to obtain
\begin{equation}\label{eq:PizPiz_thermal}
    \langle \hat{\Pi}_z^{(1)}\hat{\Pi}_z^{(2)}\rangle_{\hat{\rho}_{\text{ST}}} = \frac{1}{(\nu(T))^2} = \left(\tanh\left(\frac{ \omega}{2 T}\right)\right)^2\,.
\end{equation}
Note that the above equation does not contain any dependence on the squeezing parameter, which is consistent with the previous results \cite{Gour_position_rep}. Similarly, we can derive the Wigner representation of $\hat{\Pi}^{(i)}_x$ operator by using Eq.~\eqref{eq:Wig_trans}:
\begin{align}
    \Pi^{(i)}_x (q_i,p_i) &= \int_0^{\infty}\dd{q}\Big[e^{-2\ii p_i(q-q_i)}\delta(q-q_i)\nonumber\\
    &\phantom{={}}-e^{2\ii p_i(q+q_i)}\delta(q+q_i)\Big]\,,
\end{align}
and then use it to obtain the spin-x correlation function:
\begin{align}
    \langle \hat{\Pi}_z^{(1)}\hat{\Pi}_z^{(2)}\rangle_{\hat{\rho}} &= 2\int_0^{\infty}\!\!\!\!\dd{q_1}\dd{q_2}\!\!\int_{-\infty}^{\infty}\!\!\!\!\dd{p_1}\dd{p_2}\Big(W_{\hat{\rho}}(q_1,q_2,p_1,p_2)\nonumber\\
    &\phantom{=}-W_{\hat{\rho}}(q_1,-q_2,p_1,p_2)\Big)\,.
\end{align}
The above expression is valid for a state that respects $W_{\hat{\rho}}(q_1,q_2,p_1,p_2)=W_{\hat{\rho}}(-q_1,-q_2,p_1,p_2)$ and $W_{\hat{\rho}}(-q_1,q_2,p_1,p_2)=W_{\hat{\rho}}(q_1,-q_2,p_1,p_2)$, which is true for a TMS vacuum/thermal state, as can be seen from Eq.~\eqref{eq:Wig_TMST_appx}. Integrating over $q_i$'s and $p_i$'s, we get 
\begin{align}\label{eq:PixPix_thermal}
    \langle \hat{\Pi}_x^{(1)}\hat{\Pi}_x^{(2)}\rangle_{\hat{\rho}_{\text{ST}}} = \frac{2}{\pi}\arctan{(\sinh(2r))},
\end{align}
which, interestingly, does not contain any temperature dependence, and is exactly the same as spin-$x$ correlation function for the TMS vacuum state \eqref{eq:Bell_Gour}.  

We plug Eqs.~\eqref{eq:PizPiz_thermal} and \eqref{eq:PixPix_thermal} in Eq.~\eqref{eq:Bell_op} to obtain $\langle\hat{B}\rangle_{\hat{\rho}_{\text{ST}}}$ for the (unbinned) pseudospin operators. The dependence of $\langle\hat{B}\rangle_{\hat{\rho}_{\text{ST}}}$ on squeezing and temperature has been illustrated in Fig.~\ref{fig:binary_violation_plot_2}. 

\bibliographystyle{unsrt}
\bibliography{references}

\end{document}